\documentstyle[epsfig,12pt]{article}
\newlength{\dinwidth}
\newlength{\dinmargin}
\newcommand{\ba}{\begin{array}}
\newcommand{\ea}{\end{array}}
\newcommand{\be}{\begin{equation}}
\newcommand{\ee}{\end{equation}}
\newcommand{\bea}{\begin{eqnarray}}
\newcommand{\eea}{\end{eqnarray}}

\def\bee{\begin{eqnarray}}
\def\eee{\end{eqnarray}}
\def\be{\begin{equation}}
\def\ee{\end{equation}}

\newcommand{\beas}{\begin{eqnarray*}}
\newcommand{\eeas}{\end{eqnarray*}}

\font\cmss = cmss12
\def\half{{1 \over 2}}
\def\identity{{\rlap{1} \hskip 1.6pt \hbox{1}}}
\def\integer{{\rlap{\cmss Z} \hskip 1.8pt \hbox{\cmss Z}}}
\def\laplace{{\kern1pt\vbox{\hrule height 1.2pt\hbox{\vrule width 1.2pt\hskip
  3pt\vbox{\vskip 6pt}\hskip 3pt\vrule width 0.6pt}\hrule height 0.6pt}
  \kern1pt}}
\def\scriptlap{{\kern1pt\vbox{\hrule height 0.8pt\hbox{\vrule width 0.8pt
  \hskip2pt\vbox{\vskip 4pt}\hskip 2pt\vrule width 0.4pt}\hrule height 0.4pt}
  \kern1pt}}

\def\roughly#1{\raise.3ex\hbox{$#1$\kern-.75em\lower1ex\hbox{$\sim$}}}

\def\real{{\hbox{\cmss R} \llap{\vrule height 7.1pt width 0.4pt
depth -.1pt \hskip 0.6 pt \phantom .}}}
\def\projective{{\hbox{\cmss P} \llap{\vrule height 7.1pt width 0.4pt
depth -.1pt \hskip 0.6 pt \phantom .}}}
\def\I{{\hbox{\cmss I}}}
\def\IA{\mbox{{\hbox{\cmss IA}}}}

\def\II{\mbox{{\rlap{\cmss I} \hskip 0.7 true pt \hbox{\cmss I}}}}
\def\IIA{\mbox{{\rlap{\cmss I} \hskip 0.7 true pt \hbox{\cmss IA}}}}
\def\IIB{\mbox{{\rlap{\cmss I} \hskip 0.7 true pt \hbox{\cmss IB}}}}

\def\tIA{{$\widetilde{\hbox{\rm \IA}}$}}
\def\tRi{\widetilde{R}_1}
\def\tRii{\widetilde{R}_2}
\def\scriptRii{{\scriptstyle \tilde{R}_2}}
\def\gym{g^2_{\scriptscriptstyle YM}}
\def\gymb{\bar{g}^2_{\scriptscriptstyle YM}}
\def\tbt{{\bar{\theta} \theta}}

\begin{document}
\thispagestyle{empty}
\addtocounter{page}{-1}
\begin{flushright}
IASSNS-HEP 97-49\\
NYU-97/05/01\\
SNUTP 97-059\\
{\tt hep-th/9707099}\\
\end{flushright}
\vspace*{1.3cm}
\centerline{\Large \bf Wilson Lines and T-Duality in Heterotic
M(atrix) Theory
\footnote{
Work supported in part by NSF grant NSF-PHY-9318781, DOE grant
DE-FG02-90ER40542, a Hansmann Fellowship, the NSF-KOSEF
Bilateral Grant, KOSEF Purpose-Oriented Research Grant 94-1400-04-01-3
and SRC-Program, Ministry of Education Grant BSRI 97-2410, the Monell
Foundation and the Seoam Foundation Fellowships.}}
\vspace*{1.2cm} \centerline{\large\bf Daniel Kabat${}^{a,b}$ and
Soo-Jong Rey${}^{b,c}$}
\vspace*{0.8cm}
\centerline{\large\it Physics Department, New York University,
New York NY 10003  USA${}^a$}
\vskip0.3cm
\centerline{\large\it School of Natural Sciences, Institute for Advanced
Study}
\vskip0.1cm
\centerline{\large\it Olden Lane, Princeton NJ 08540 USA${}^b$}
\vskip0.3cm
\centerline{\large\it Physics Department, Seoul National University,
Seoul 151-742 KOREA${}^c$}
\vspace*{0.6cm}
\centerline{\large\tt kabat@sns.ias.edu, sjrey@gravity.snu.ac.kr}
\vspace*{1.5cm}
\centerline{\large\bf abstract}
\vskip0.5cm
We study the M(atrix) theory which describes the $E_8 \times E_8$
heterotic string compactified on ${\bf S}^1$, or equivalently M-theory
compactified on an orbifold $({\bf S}^1/\integer_2) \times {\bf S}^1$,
in the presence of a Wilson line.  We formulate the corresponding
M(atrix) gauge theory, which lives on a dual orbifold ${\bf S}^1
\times ({\bf S}^1 / \integer_2)$.  Thirty-two real chiral fermions
must be introduced to cancel gauge anomalies.  In the absence of an
$E_8 \times E_8$ Wilson line, these fermions are symmetrically
localized on the orbifold boundaries.  Turning on the Wilson line
moves these fermions into the interior of the orbifold.  The M(atrix)
theory action is uniquely determined by gauge and supersymmetry
anomaly cancellation in 2+1 dimensions.  The action consistently
incorporates the massive \IIA~supergravity background into M(atrix)
theory by explicitly breaking (2+1)-dimensional Poincar\'e invariance.
The BPS excitations of M(atrix) theory are identified and compared to
the heterotic string.  We find that heterotic T-duality is realized as
electric-magnetic S-duality in M(atrix) theory.

\vspace*{1.1cm}

\setlength{\baselineskip}{18pt}
\setlength{\parskip}{12pt}

\newpage
\section{Introduction}

In the strong coupling limit all known superstring theories are
unified into eleven-dimensional M-theory \cite{witten}.  But little
was known about the fundamental constituents of M-theory, until Banks
{\sl et.~al.}~\cite{bfss} proposed a beautiful partonic definition of
M-theory (for earlier hints, see \cite{townsend}).  They argued that
the partons can be identified from the strongly coupled type
\IIA~string by boosting it infinitely along the eleventh `quantum'
direction. The light-front view in this limit is of infinitely many
zero-branes threaded on the strongly coupled type \IIA~string itself.
One thus discovers that the fundamental degrees of freedom of M-theory
consist of zero-branes and the infinitely short open strings gluing
them together.  This suggests that M-theory parton dynamics are
governed by the large N limit of ${\cal N} = 16$ supersymmetric, gauge
group U(N), matrix quantum mechanics~\cite{matrixqm}.

While M-theory arises as the strong coupling limit of any perturbative
superstring theory, the M-theory partons and their dynamics are most
easily identified from the type \IIA~string.  Starting from the
heterotic string, on the other hand, identifying the M-theory partons
appears obscure.  It has been shown~\cite{horavawitten} that
the $E_8 \times E_8$ heterotic string is related to M--theory via
compactification on an ${\bf S}^1/\integer_2$ orbifold of the eleventh
quantum direction.  On the orbifold, there are no propagating
Kaluza-Klein excitations, only standing waves.  So it is not possible
to boost the heterotic string along the quantum orbifold direction.

Given this kinematical limitation, one must boost the heterotic string
along a classical direction to go to the infinite momentum frame.
This can be understood using heterotic --- type \IA~S-duality~\cite{
polchinskiwitten}, which interchanges the quantum orbifold direction
with one of the classical non-compact directions, and lets one
formulate heterotic M(atrix) theory as a $\integer_2$ orbifold of the
original type \IIA~M(atrix) theory~\cite{motl, kimrey, banksmotl}.
This heterotic M(atrix) theory is the large N limit of the ${\cal N} =
8$ supersymmetric $O(N)$ matrix quantum mechanics studied
in~\cite{danielssonferretti, kachrusilverstein, lowe,
banksseibergsilverstein}, and has been further developed
in~\cite{loweII, sjrey, Horava}.  It differs from the original type
\IIA~M(atrix) theory both by the choice of gauge group and by the
presence of a twisted sector: at each orbifold fixed point there are
16 real fermions in the fundamental representation of the $O(N)$ gauge
group.

We are interested in compactifying the heterotic theory on an
additional ${\bf S}^1$.  Following the prescription of \cite{taylor,
banksmotl}, the resulting theory is most easily described as a
(2+1)-dimensional gauge theory on an orbifold ${\bf S}^1 \times ({\bf
S}^1 / \integer_2)$.  In section 2 we construct this theory as an
orbifold of a gauge theory on ${\bf T}^2$. The main purpose of this
paper is then to study the M(atrix) theory when a Wilson line is turned
on in the heterotic theory.  By imposing gauge and supersymmetry anomaly
cancellation, we formulate M(atrix) theory in the presence of a
Wilson line in section 3.  In section 4, we study the BPS spectrum of
the M(atrix) theory, and compare it to the heterotic string.  A highly
non-trivial test is that the M(atrix) theory should reproduce the
T-duality of the perturbative heterotic string.  We find that
heterotic T-duality is realized as electric-magnetic S-duality in
M(atrix) theory, reminiscent of the way T-duality is realized in type
\IIA~\cite{susskind, ganorramgoolamtaylor}.

\section {Heterotic M(atrix) Theory for Unbroken $E_8 \times E_8$}

M-theory compactified on a cylinder $({\bf S}^1/\integer_2) \times
{\bf S}^1$ provides a unified description of the heterotic and type
\IA~string theories~\cite{horavawitten}.  In this section, we
formulate the M(atrix) theory description of this compactification,
assuming that $E_8 \times E_8$ gauge symmetry is unbroken.  Toroidal
compactification of heterotic M(atrix) theory has been discussed by
Banks and Motl~\cite{banksmotl}, and following their work, we construct
the
M(atrix) theory as a $\integer_2$ orbifold of a gauge theory on ${\bf
T}^2$.

Our notations are as follows. In M-theory, we fix the eleven
dimensional Planck length $\ell_{11}$ and compactify on ${\bf T}^2 =
{\bf S}^1 \times {\bf S}^1$.  The circles are along the $x^1$- and
$x^2$-directions, with radii $R_1$ and $R_2$, respectively.  The
$x^{11}$-direction along which M-theory is boosted is compactified on
a regulator circle with radius $R_{11} \rightarrow \infty$.

The heterotic compactification of M-theory is obtained as a
$\integer_2$ orbifold by $x^1 \rightarrow - x^1$.  This describes the
heterotic string compactified on the circle in the $x^2$-direction.  The
heterotic string parameters are
\be
\label{HetParams}
\hbox{\rm heterotic} \enskip : \hskip0.5cm
g_{\rm H} = \left({R_1 \over \ell_{11}}\right)^{3/2}
\hskip1cm \alpha'_{\rm H} = {\ell_{11}^3 \over R_1}
\hskip1.2cm
R_{\rm H} = R_2.
\ee
By instead taking $x^{11}$ to be the `quantum' direction, this can also
be regarded as a type \IA~compactification of M-theory, that is, a
compactification on $({\bf S}^1/\integer_2) \times {\bf S}^1$, where
the interval has length $\pi R_1$ and the circle has radius $R_2$.
The type \IA~parameters are given by
\[
\hbox{\cmss IA} \enskip : \hskip0.5cm g_A =
\left({R_{11} \over \ell_{11}}\right)^{3/2}
\hskip1cm \alpha'_A = { \ell_{11}^3 \over R_{11}}.
\]

A total of 16 D8-branes are present in type \IA, with a gauge field
$B_\mu$ propagating on their worldvolumes.  The moduli of type
\IA~include the positions of the D8-branes along the $x^1$ orbifold
direction, as well as the Wilson line $B_2$ around the $x^2$ circle
direction.  We wish to go to the point in moduli space where the
8-brane gauge symmetry is enhanced to $E_8 \times E_8$ in the limit of
infinite type \IA~coupling.  This corresponds to a symmetric
configuration of Wilson lines, in which 8 D8-branes are located at
each orientifold and a Wilson line $B_2 = \left(0^8,
({\scriptscriptstyle 1 \over 2})^8\right)$ is turned on.

To show that this is correct, we apply T-duality to the $x^1$ orbifold
direction, to go to \mbox{type \I} compactified on a torus.  T-duality
maps D-brane positions into Wilson lines~\cite{PolchinskiReview}, so
in type \I~there are Wilson lines $B_1 = B_2 = \left(0^8,
({\scriptscriptstyle 1 \over 2})^8\right)$ turned on.  Following
\cite{kachrusilverstein}, one can then use S-duality to show that the
correct $E_8 \times E_8$ multiplets appear in the limit of infinite
type \IA~coupling.

We then apply T-duality to this system for a second time,
along the $x^2$-direction.
This takes us to a new type \IA~theory, which we refer to as type
\tIA.  The \tIA~parameters are
\be
\label{tIAParams}
\widetilde{\hbox{\rm \IA}} \enskip : \hskip0.5cm
{\widetilde g}_A = {\ell_{11}^{3/2} R_{11}^{1/2} \over R_1 R_2}
\hskip1cm
\tRi = {\ell_{11}^3  \over R_{11} R_1}
\hskip1cm \tRii = {\ell_{11}^3 \over R_{11} R_2}.
\ee
It is important to note that this T-duality has interchanged the
orbifold and circle directions.   That is, type \tIA~is compactified on
${\bf S}^1 \times ({\bf S}^1/\integer_2)$, where now $x$ denotes the
circle direction with radius $\tRi$ and $y$ denotes the orbifold
direction with length $\pi \tRii$.  In the symmetric configuration
corresponding to unbroken $E_8 \times E_8$, a
Wilson line
$B_1 = \left(0^8, ({\scriptscriptstyle 1 \over 2})^8\right)$ is turned on
and there are 8 D8-branes at each orientifold.  See Fig.~1.

Note that a Wilson line deformation in the heterotic string
corresponds to a deformation of the Wilson line
$B_2$ in type \IA.  After T-dualizing
to type \tIA~this maps to a deformation of the positions of the
D8-branes.  That is, when a heterotic Wilson line is turned on, the
D8-branes in type \tIA~are no longer symmetrically located at the two
orientifolds, but rather move into the bulk of the orbifold. See Fig.~2.

\begin{figure}
\epsfig{file=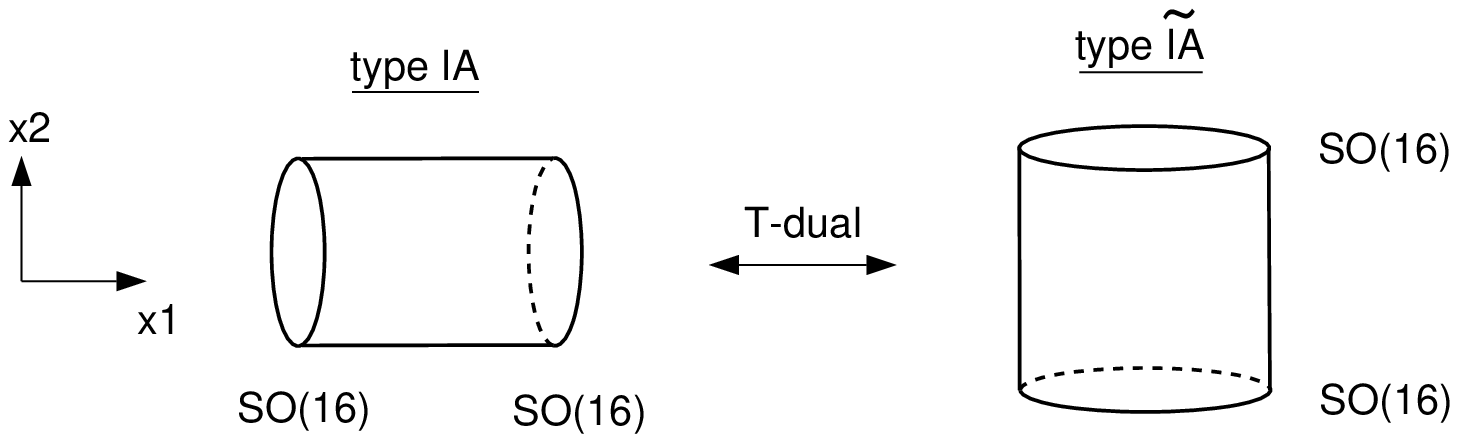}
\caption{Type \IA~and \tIA~compactifications for unbroken $E_8 \times
E_8$.  There are 8 D8-branes at every orientifold, giving $SO(16)$ gauge
symmetry.}
\vspace{1.8 cm}
\epsfig{file=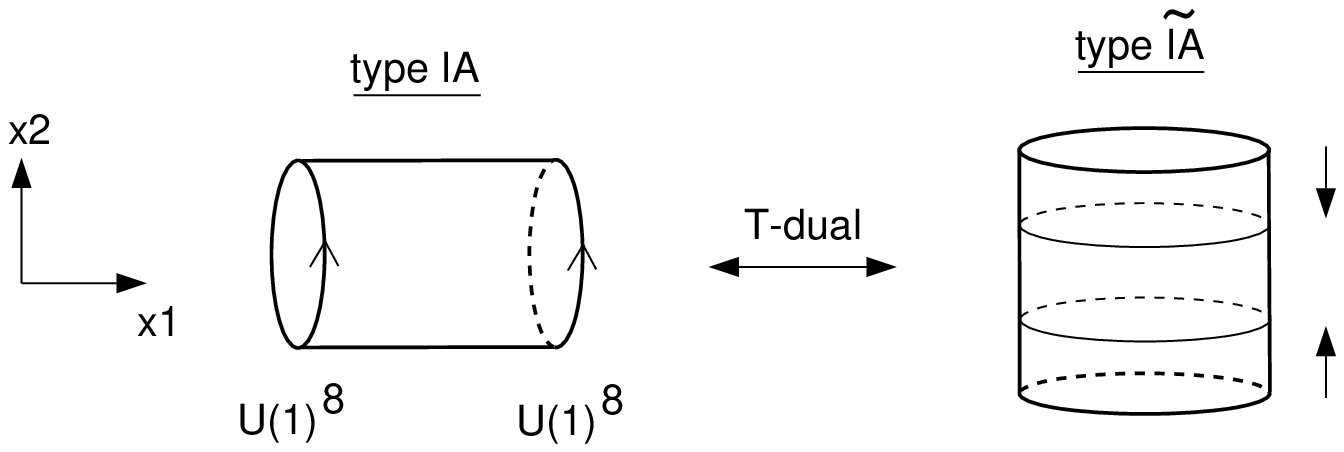}
\caption{Turning on a heterotic Wilson line deforms the Wilson line
$B_2$ in type \IA, breaking $SO(16)$ to $U(1)^8$.  In type \tIA~it
moves the D8-branes away from the orientifolds.}
\end{figure}

To construct the heterotic M(atrix) theory which describes this
compactification, we start from type \IIA~compactified on ${\bf T}^2$.
This can be described in M(atrix) theory as a 2+1 dimensional $U(N)$
gauge theory with 16 supercharges, compactified on the T-dual torus
$\widetilde{{\bf T}}^2$~\cite{taylor}.  This theory has a $\integer_2$
symmetry, which we mod out by to obtain the untwisted sector of the
heterotic M(atrix) theory on ${\bf S}^1 \times ({\bf S}^1 /
\integer_2)$.  A twisted sector, localized at the orbifold fixed
points, must then be introduced for gauge anomaly cancellation.
Following \cite{banksmotl}, we carry out this construction in the next
two subsections.

\subsection{Type \IIA~M(atrix) Theory}

We begin by reviewing the description of the type \IIA~M(atrix) theory
on ${\bf T}^2$.  This is given by a 2+1 dimensional gauge theory with
gauge group $U(N)$ and 16 supercharges, compactified on the dual torus
$\widetilde{{\bf T}}^2$ \cite{taylor}.  The appropriate gauge theory
can be obtained by dimensional reduction of ten-dimensional
supersymmetric $U(N)$ Yang-Mills theory.  Thus the M(atrix) gauge
theory has an $Spin(7)_R$ R-symmetry, which may be viewed as originating
from the reduced seven dimensions.

Our spinor conventions are as follows.  We denote $SO(2,1)$ vector
indices by $\alpha, \beta = 0,1,2$ and $SO(2,1)$ spinor indices by
$a,b=1,2$.  The (2+1)-dimensional Dirac matrices
$\left(\gamma^\alpha\right)_{ab}$ are $2 \times 2$ purely imaginary
matrices:
\[
\gamma^0 = \sigma^2 = \left( \begin{array}{cc} 0 & - i \\ + i & 0
\end{array} \right),
\hskip0.5cm
\gamma^1 = i \sigma^1 = \left( \begin{array}{cc} 0 & + i \\
+ i & 0 \end{array} \right),
\hskip0.5cm
\gamma^2 = i \sigma^3 = \left( \begin{array}{cc}
+ i & 0 \\ 0 & - i \end{array} \right).
\]
These obey $\lbrace \gamma^\alpha, \gamma^\beta \rbrace = - 2
\eta^{\alpha \beta} \identity$ with metric $\eta_{\alpha\beta} =
(-++)$.  We define $\gamma^{\alpha\beta} = \half
[\gamma^\alpha,\gamma^\beta]$.  In odd dimensions, there are two
inequivalent representations of the Clifford algebra (both of which
give identical representations of the Lorentz group).  These matrices
have been chosen to satisfy $\gamma^{(3)} \equiv i \gamma^0 \gamma^1
\gamma^2 = + \identity$.  An inequivalent representation of the
Clifford algebra is provided by changing the sign of all
$\gamma^\alpha$.  We also introduce a set of real, antisymmetric
$Spin(7)_R$ Dirac matrices $\left(\gamma^i\right)_{AB}$,
where $i,j = 3,\cdots,9$ are $Spin(7)_R$ vector indices and $A,B =
1,\cdots,8$ are $Spin(7)_R$ spinor indices.  These matrices obey
$\lbrace \gamma^i, \gamma^j \rbrace = - 2 \delta^{ij}\identity$,
and we take
them to satisfy $\gamma^{(8)} \equiv \gamma^3 \cdots \gamma^9 = +
\identity$.  The ten-dimensional Dirac matrices can then be
constructed as a tensor product.
\[
\Gamma^\alpha = \left[ \begin{array}{cc}
0 & \gamma^\alpha \otimes \identity_{8 \times 8} \\
\gamma^\alpha \otimes \identity_{8 \times 8} & 0 \end{array} \right]
\hskip1cm
\Gamma^i = \left[ \begin{array}{cc}
0 & -i \identity_{2 \times 2} \otimes \gamma^i \\
+ i \identity_{2 \times 2} \otimes \gamma^i & 0 \end{array} \right]
\]
This is in a Majorana-Weyl representation, with
\[
\Gamma^{(11)} \equiv \Gamma^0 \cdots \Gamma^9 = \left[
\begin{array}{cc}
+ \identity & 0 \\
0 & - \identity
\end{array} \right].
\]

The field content of the M(atrix) gauge theory is a gauge field
$A_\alpha$, seven adjoint scalar fields $X_i$, and an adjoint spinor
field $\psi_{aA}$ satisfying $\Gamma^{(11)} \psi = \psi$.  The
Lagrangian, obtained by dimensional reduction from (9+1) dimensions,
is given by
\bea
L & = & - {1 \over 2 \gym} \int_{{\widetilde T}_2} d^2x \, {\rm Tr}
\Big\lbrace F_{\alpha \beta} F^{\alpha \beta}
+ 2 D_\alpha X_i D^\alpha X^i - [X_i, X_j][X^i,X^j] \nonumber \\
& & \qquad\qquad\qquad\qquad
 - 2 i {\overline \psi}_A \gamma^\alpha D_\alpha \psi_{A}
+ 2 i {\overline \psi}_A \gamma^i_{AB} [X_i, \psi_B] \,\, \Big\rbrace
\label{untwisted}
\eea
where we have suppressed the $SO(2,1)$ spinor indices.  This action
can be regarded as the world-volume action for a stack of $N$
D2-branes wrapped on ${\widetilde T}_2$.  The gauge coupling is
related to the M-theory and type \tIA~parameters by
\bea
\label{YMcoupling}
\gym & = & {2 R_{11} \over R_1 R_2} \nonumber \\
     & = & 2 \widetilde{g}_A / \sqrt{\alpha'_A}
\eea
as can be seen by normalizing the energy of a quantum of electric flux
to agree with the energy of a D0-brane Kaluza-Klein mode.

This action is invariant under the `dynamical' supersymmetry
transformations
\bee
\delta_\epsilon A_\alpha &=& + {i \over 2} \, {\overline \epsilon}_A
\gamma_\alpha \psi_A \nonumber \\
\delta_\epsilon X^i &=&
-{1 \over 2} \, {\overline \epsilon}_A \gamma^i_{AB} \psi_B \nonumber \\
\delta_\epsilon \psi_A
&=& - {1 \over 4} F_{\alpha \beta} \gamma^{\alpha \beta} \epsilon_A
- {i \over 2} D_\alpha X_i \gamma^\alpha \gamma^i_{AB} \epsilon_B
- {i \over 4} [X_i, X_j] \gamma^{ij}_{AB} \epsilon_B \,.
\label{DynSusy}
\eee
It is also invariant under the `kinematical' supersymmetry transformations
\bee
\delta_\eta A_\alpha &=& \delta_\eta X_i = 0
\nonumber \\
\delta_\eta \psi_A &=& \eta_A \identity.
\label{fermionicsymm}
\eee
Note that kinematical supersymmetry acts only on the center-of-mass
$U(1)$ part of the $U(N)$ M(atrix) gauge group.

The M-theory origin of these symmetries is easily understood
\cite{banksseibergshenker}.  M-theory has a total of 32
supersymmetries, which decompose into ${\bf 16}_+ \oplus {\bf 16}_-$
of $SO(9,1)$.  The choice of infinite momentum frame breaks the
supersymmetries in the ${\bf 16}_-$, which become the
non-linearly realized kinematical supersymmetries of M(atrix) theory.
The supersymmetries in the ${\bf 16}_+$ are unbroken and give rise to
the dynamical supersymmetries of M(atrix) theory.  Further decomposing
under $SO(2,1) \times SO(7) \subset SO(9,1)$, note that the
2+1-dimensional spinors $\epsilon_A$ and $\eta_A$ should be taken to
be in inequivalent (opposite-sign) representations of the
(2+1)-dimensional Clifford algebra.

\subsection{Heterotic M(atrix) Theory}

The \IIA~M(atrix) theory Lagrangian is invariant under a
combined $\integer_2$ operation $\Omega \cdot P$, where $\Omega$
corresponds to orientation reversal in string theory, and $P$ is a
(2+1)-dimensional parity transformation.  Orientation reversal
acts as
\bea
\label{OmegaAction}
\Omega \hskip 0.5 cm : \hskip 0.5 cm
A_\alpha(x, y) &\rightarrow& \pm \, A_\alpha^{\rm T} (x, y)
\nonumber \\
X_i (x, y) &\rightarrow& \pm \, X_i^{\rm T} (x, y) \\
\psi_{aA} (x, y) &\rightarrow& \pm \, \psi_{aA}^{\rm T} (x, y) \,\,.
\nonumber
\eea
We will only consider the $(-)$ choice of sign in the definition of
$\Omega$, for reasons given below.  The (2+1)-dimensional parity
transformation $P$ acts as
\bea
\label{PAction}
P \hskip 0.5 cm : \hskip 0.5 cm
A_{0,1} (x, y) &\rightarrow& + \, A_{0,1} (x, - y)
\nonumber \\
A_2 (x, y) &\rightarrow& - \, A_2 (x, - y) \nonumber \\
X_i (x, y) &\rightarrow& - \, X_i (x, - y) \\
\psi_{aA} (x, y) &\rightarrow&
+ i (\gamma^2)_{ab} \, \psi_{bA} (x, - y) \,\,. \nonumber
\eea
Note that the scalar fields $X_i$ are taken to be pseudo-scalar.

As in~\cite{banksmotl}, heterotic M(atrix) theory is obtained from
type \IIA~M(atrix) theory
by modding out by $\Omega \cdot P$. In other words, the heterotic
M(atrix) theory is defined as a parameter space orbifold of the
\IIA~M(atrix) theory\footnote{This construction generalizes
straightforwardly to higher-dimensional orbifold compactifications for
which nontrivial orbifold boundaries arise, as studied
in~\cite{fayyasmith, kimrey2}.}.
The parity transformation $P$
acts on the parameter space as an involution, so the dual torus
$\widetilde{{\bf T}}^2$ becomes a dual orbifold with cylinder
topology:
\beas
{\widetilde C}_2 & = & {\bf S}^1 \times ({\bf S}^1/\integer_2) \\
\noalign{\vskip 0.2 cm}
x & \approx & x + 2 \pi {\widetilde R}_1 \\
0 & \le & y \,\,\, \le \,\,\, \pi {\widetilde R}_2 \,\,\, .
\eeas
In addition, at the orbifold fixed circles $y = 0$ and $y = \pi
{\widetilde R}_2$, the action of $\Omega \cdot P$ imposes boundary
conditions on the fields:
\bee
(A_0, \,\, A_1, \,\, \psi_{2A}) (x)
& & {\tt antisymmetric}
\nonumber \\
(A_2, \,\,X^i, \,\,\psi_{1A})(x)
& & {\tt symmetric}
\nonumber \\
\partial_2 (A_0, \,\, A_1, \,\,  \psi_{2A}) (x)
& & {\tt symmetric}
\nonumber \\
\partial_2 (A_2, \,\, X^i, \,\, \psi_{1A})(x)
& & {\tt antisymmetric} \,\,\, .
\label{bc}
\eee
Note that these conditions imply that $F_{01}$ is antisymmetric on the
boundary, while $F_{02}$ and $F_{12}$ are symmetric.  In particular,
the $U(1)$ part of the electric field in the circle direction
vanishes identically on the boundary, which will be important later in
establishing S-duality of M(atrix) theory.  Heuristically, this may
be understood as follows.  A propagating photon has an electric field
which is orthogonal to its direction of motion.  Along the orbifold
direction, there is no propagating photon.  Consequently, along the
circle direction, there is no electric field.

These boundary conditions modify the M(atrix) gauge theory in several
ways.  First, they break half the supersymmetry.  The dynamical
supersymmetry parameters $\epsilon_A$ appearing in (\ref{DynSusy})
must be taken to be invariant under the $\integer_2$ projection,
$\epsilon_A = i \gamma^2 \epsilon_A$.  We will refer to this amount of
supersymmetry as ${\cal N} = (0,8)$ supersymmetry in 2+1 dimensions.
At the same time, to respect the boundary conditions on the fermions,
the kinematical supersymmetry parameters $\eta_A$ must satisfy $\eta_A
= - i \gamma^2 \eta_A$.  This sign difference reflects the fact that
$\epsilon_A$ and $\eta_A$ are in inequivalent representations of the
(2+1)-dimensional Clifford algebra.

Second, at the boundaries, only gauge transformations in an $O(N)$
subgroup of $U(N)$ are allowed, since a gauge transformation $A_\alpha
\rightarrow U \left(A_\alpha - i \partial_\alpha \right)U^{-1}$
respects the boundary conditions only if $U^{\rm T} U \vert_{y=0,
\pi \tRii} = \identity$.  Note that under the boundary $O(N)$ gauge
group $\lbrace A_0, A_1, \psi_{2A} \rbrace$ transform in the adjoint
representation, while $\lbrace A_2, X^i, \psi_{1A} \rbrace$ are in the
symmetric representation.

Finally, but most importantly, the boundary conditions lead to the
gauge anomaly discussed in \cite{kimrey, banksseibergsilverstein}.
The fermions $\psi_A$ have normalizeable modes which are independent
of the coordinate $y$.  These modes behave as if they were fermions
in 1+1 dimensions.  From the 2+1-dimensional Dirac equation, we find
that the upper components $\psi_{1A}$ are right-moving, while the
lower components $\psi_{2A}$ are left-moving.  These modes therefore
generate a left-moving
$O(N)$ gauge anomaly proportional to $8 I_2({\tt adj.})  -
8 I_2({\tt symm.}) = - 32 I_2({\tt fund.})$.  Here $I_2({\tt R})$ is
the quadratic index of the representation ${\tt R}$.  An argument given
in \cite{horavawitten}~shows that half of the
anomaly must be symmetrically localized on each orbifold boundary.

The anomaly can be cancelled by introducing 32 left-moving
Majorana-Weyl fermions in the fundamental representation of $O(N)$.
For sufficiently large $N$, this is the unique choice of
representation which cancels the anomaly.  In order to cancel the
anomaly {\sl locally}, half of these fermions must be localized at
each orbifold boundary.  They can be regarded as a twisted sector of
the M(atrix) theory $\integer_2$ orbifold.  From the string theory
point of view, they can be understood as coming from the 2-8 strings
present in type \tIA~string theory~\cite{kachrusilverstein}.

Note that, had we chosen the $(+)$ sign in the definition of $\Omega$,
the boundary gauge group would be $USp(N)$~\cite{kimrey}.
In this case, there is no
way to cancel the gauge anomaly, since negative numbers of fermions
would have to be introduced.

To summarize, the untwisted sector of heterotic M(atrix) theory
is described by a (2+1)-dimensional gauge theory on the dual
orbifold ${\tilde C}_2$:
\bee
L_{\rm untwisted} & = &  - {1 \over \gym} \int_{{\widetilde C}_2} d^2x \,
{\rm Tr} \Big\lbrace F_{\alpha \beta} F^{\alpha \beta}
+ 2 D_\alpha X_i D^\alpha X^i - [X_i, X_j][X^i,X^j] \nonumber \\
& & \qquad\qquad\qquad\qquad
 - 2 i {\overline \psi}_A \gamma^\alpha D_\alpha \psi_{A}
+ 2 i {\overline \psi}_A \gamma^i_{AB} [X_i, \psi_B] \,\, \Big\rbrace
\label{HetUntwisted}
\eee
An overall factor of two has been inserted because we write the action
only on the fundamental domain of the $\integer_2$ action.

The twisted sector involves real chiral fermions $\chi^{(1)}$ and
$\chi^{(2)}$ localized at $y = 0$ and $\pi \tRii$, respectively.
They are left-moving, so these fermions are $(0,8)$ supersymmetry
singlets.  They are in the fundamental representation of the boundary
M(atrix) gauge group $O(N) \subset U(N)$, and under the $SO(16) \times
SO(16)$ symmetry associated with the 8 D8-branes present at each
orientifold, they transform as $({\bf 16}, {\bf 1})$ and $({\bf 1},
{\bf 16})$, respectively.  Their Lagrangian is ($\partial_\pm =
\partial_0 \pm \partial_1$)
\bee
L_{\rm twisted} & = & i \oint d x \,  \Big\lbrace \,\,
\chi^{(1)} \left(\partial_- + i \left.A_-\right\vert_{y = 0}
+ i B^{(1)}_- \right) \chi^{(1)}
\nonumber \\
& & \qquad\qquad\!\!\! + \chi^{(2)} \left(
\partial_- + i \left.A_-\right\vert_{y = \pi \tRii}
+ i B^{(2)}_- \right) \chi^{(2)}\,\, \Big\rbrace.
\label{twisted}
\eee

We have included couplings to the background fields $B^{(1)}$ and
$B^{(2)}$, which are the $SO(16)$ gauge fields that propagate on the
D8-branes at $y = 0$ and $y = \pi \tRii$.  Recall that the $E_8 \times
E_8$ heterotic string corresponds to Wilson lines
$\left(B^{(1)},B^{(2)}\right) = \left(0^8,({\scriptscriptstyle 1 \over
2})^8 \right)$.  The (1+1)-dimensional dynamics of the gauge fields
$(A_0, A_1)$ is not explicitly given but is tacitly assumed to be part
of the untwisted sector Lagrangian (\ref{HetUntwisted}).

The twisted sector does not seem to have a direct derivation from the
underlying \IIA~M(atrix) theory, and must be introduced by hand (see
however \cite{Horava}).  But it is necessary for internal consistency.
Note that the boundary conditions on $A_0$ and $A_1$ force them to be
purely imaginary and antisymmetric at the orbifold fixed points, as
appropriate for them to couple to the (1+1)-dimensional real chiral
fermions located at the boundaries.

\section{Turning on Wilson lines}

As we discussed in the previous section, the heterotic string
compactified on a circle is equivalent to the strong-coupling limit of
type \IA~on an $({\bf S}^1/\integer_2) \times {\bf S}^1$ orbifold.
Unbroken $E_8 \times E_8$ gauge symmetry corresponds to a symmetric
configuration, in which 8 D8-branes (plus their images) are located at
each $\integer_2$ fixed point and a Wilson line $B_2 =
\left(0^8,({\scriptscriptstyle 1 \over 2})^8\right)$ is turned on
around the ${\bf S}^1$.

We now wish to generalize this, by turning on a Wilson line in the
heterotic string.  Taking the Wilson line to lie in an $SO(16) \times
SO(16)$ subgroup of $E_8 \times E_8$, it is easy to see what this
corresponds to in terms of the equivalent type \IA~theory.  The eight
coincident D8-branes at each end of the cylinder have a world-volume
gauge group $SO(16) \times SO(16)$.  Turning on the heterotic Wilson
line corresponds to deforming the Wilson line $B_2$ in the world-volume
theory of these D8-branes.

To describe this in M(atrix) theory, we apply T-duality to both
directions of the orbifold $({\bf S}^1/\integer_2) \times {\bf S}^1$.
This takes us to type \tIA~theory on the dual orbifold ${\bf S}^1
\times ({\bf S}^1/\integer_2)$.  Again, we emphasize that the duality
interchanges the orbifold and circle directions, and maps the Wilson
line $B_2$ in type \IA~to the positions of the D8-branes along the
${\bf S}^1/\integer_2$ in type \tIA.  So for unbroken $E_8 \times
E_8$, type \tIA~will have a Wilson line
$B_1 = \left(0^8,\left({\scriptscriptstyle 1 \over 2}\right)^8\right)$
turned on and 8 D8-branes located at each orientifold.
Turning on a heterotic Wilson line
moves the D8-branes away from the type \tIA~orientifolds, while
leaving the \tIA~Wilson line $B_1$ unchanged.
See Figs. 1 and 2.

The D8-branes give rise to chiral fermions in the M(atrix) theory,
which can be thought of as 2-8 strings in type \tIA~string theory.
When the 8-branes move away from the ends of the cylinder, these 2-8
strings will move with them.  So we expect that the M(atrix) theory
description of the heterotic string with a Wilson line will involve a
(2+1)-dimensional Yang-Mills theory on ${\bf S}^1 \times ({\bf
S}^1/\integer_2)$, coupled to 16 complex chiral fermions $\chi_I$ which
propagate on circles at fixed positions $y = y_I$, $I=1,\ldots,16$
along the orbifold direction\footnote{This interpretation of a heterotic
Wilson line has also been considered by L.~Motl.  We thank T.~Banks for
informing us of this.}.

This raises an interesting puzzle.  From the spacetime point of view,
the D8-branes are sources of R-R and NS-NS fields.  Tadpoles from disc
diagrams localized near the D8-branes produce a flux that is absorbed
by $\real \projective^2$ diagrams localized near the orientifolds.
When the D8-branes are symmetrically distributed at the ends of the
cylinder, this tadpole cancellation takes place locally, and the
spacetime fields in the bulk of the cylinder are constant.  But when
the D8-branes move away from the type \tIA~orbifold boundaries, they
create a flux along the axis of the cylinder.  The resulting R-R and
NS-NS fields were found in \cite{polchinskiwitten}~by solving
spacetime equations of motion.  How does the M(atrix) gauge theory
reproduce this physics?

A key observation is that the formulation of the M(atrix) theory we
sketched above cannot be the complete story --- it has a gauge
anomaly.  The fermions $\psi_A$ have modes, constant along the
orbifold direction, which are chiral in the (1+1)-dimensional sense.
These modes give rise to a gauge anomaly, which is symmetrically
distributed between the two ends of the cylinder.  The fermions
$\chi_I$, which can be thought of as 2-8 strings, had to be introduced
to cancel the anomaly.  When these fermions are symmetrically
distributed between the two ends of the cylinder the anomaly
cancellation is local.  But when the D8-branes move away from the
boundaries of the \tIA~orbifold, the $\chi_I$ fermions move with them,
and anomaly cancellation is possible only if there is an additional
interaction present in the M(atrix) gauge theory which can `move the
anomaly' from the positions of the D8-branes to the ends of the
cylinder.

This leads us to state that the correct formulation of M(atrix) theory
in the presence of a Wilson line is to be found by enforcing internal
consistency conditions on the (2+1)-dimensional Yang-Mills theory,
namely cancellation of gauge and supersymmetry anomalies.  For the
${\bf S}^1$ compactification that we are considering, this is
sufficient to uniquely determine the M(atrix) theory Lagrangian, at
least through two-derivative order\footnote{It is unclear how to
formulate M(atrix) theory in compactifications with less
supersymmetry.}.  As we shall see, an internally consistent
(2+1)-dimensional
action will automatically correctly take into account both
the R-R and NS-NS background fields produced by the D8-branes.  It
achieves this by explicitly breaking (2+1)-dimensional
Poincar\'e invariance.

This provides an interesting variation on the idea of D-branes as
probes of background geometry~\cite{DouglasProbes,
BanksDouglasSeiberg, Seiberg3d, Seiberg5d}.  The configurations
studied previously have involved systems of parallel branes, so the
field theory living on the probe is Poincar\'e invariant.  Typically
the probe theory has a flat classical moduli space.  But one finds
that quantum corrections to the moduli space metric precisely encode
the non-trivial spacetime geometry that is established by the other
branes.

In the system we are considering, which can be thought of as an
intersecting 2-brane -- 8-brane system in type \IIA, it is likewise
true that the field theory on the probe 2-brane knows about the
background geometry due to the 8-brane.  But it is now encoded into
(non-Poincar\'e-invariant) terms which must be added to the classical
probe action, in order to obtain a consistent (supersymmetric and
anomaly-free) probe theory.

\pagebreak[3]
\subsection{Gauge Anomaly Cancellation\protect\footnotemark}
\footnotetext{After obtaining the results in this section we
received a paper by Ho\u{r}ava \cite{Horava}~discussing this
mechanism.}

When a heterotic Wilson line is turned on, the D8-branes of type
\tIA~move away from the orbifold boundaries, to positions $y = y_I$,
$I=1,\ldots,16$ along the orbifold direction.  So we expect the
M(atrix) theory to include 16 (1+1)-dimensional complex chiral
fermions $\chi_I$, which can be thought of as unexcited 2-8 strings,
localized at positions $y_I$ in the bulk of the orbifold.

These chiral fermions are in the fundamental representation of the
M(atrix) theory $U(N)$ gauge group, and should couple in the usual way
($\partial_\pm = \partial_0 \pm \partial_1$).
\be
\label{ChiAction}
S_\chi = i \int d^3x \sum_{I=1}^{16} \delta(y - y_I) \overline{\chi}_I
\left(\partial_- + i A_- + i B_-^{(I)} \right) \chi_I
\ee
Here $A$ is the $U(N)$ gauge field, and $B^{(I)}$ is the $U(1)$ gauge
field on the $I^{th}$ D8-brane.  The effective action $\Gamma_I[A]$
obtained by integrating out the the field $\chi_I$ is not gauge
invariant.  Rather, it has an anomalous variation under gauge
transformations $\delta_\Lambda A = d\Lambda + i [A,\Lambda]$.
\be
\label{ChiAnomaly}
\delta_\Lambda \Gamma_I = - {1 \over 4 \pi} I_2({\tt fund.})
\int {\rm Tr} \left. \left( \Lambda dA\right)\right\vert_{y = y_I}
\ee

There is no gauge anomaly in odd dimensions, so the (2+1) dimensional
Yang-Mills theory must be anomaly-free in the bulk of the dual
orbifold.  But, as we discussed in the previous section, the fermions
$\psi_A$ have normalizeable modes which are independent of $y$.  One
can think of these modes as if they were fermions in 1+1 dimensions.
The upper component $\psi_{1A}$ is a right-moving symmetric tensor,
while the lower component $\psi_{2A}$ is a left-moving adjoint.  So
integrating out these modes induces an effective action $\Gamma_\psi$
with a gauge anomaly.  As in \cite{horavawitten}, the anomaly must be
a sum of two terms.  Each term is localized at one end of the
cylinder, and has half the strength of the standard 1+1 dimensional
anomaly.  Compared to $\Gamma_I$, an additional factor of $1/2$ arises
because the fermions $\psi_A$ are real instead of complex.
\bea
\delta_\Lambda \Gamma_\psi & = & - {1 \over 16 \pi} \left(
8 I_2({\tt adj.}) - 8 I_2({\tt symm.})\right) \int {\rm Tr}
\left\lbrace\Lambda dA\vert_{y=0} + \Lambda dA \vert_{y = \pi \tRii}
\right\rbrace \nonumber \\
& = & 8 \, {1 \over 4 \pi} I_2({\tt fund.}) \int {\rm Tr} \left\lbrace
\Lambda dA \vert_{y = 0} + \Lambda dA \vert_{y = \pi \tRii}
\right\rbrace\label{PsiAnomaly}
\eea
Here $I_2({\tt adj.}) = N-2$ and $I_2({\tt symm.}) = N+2$ are the quadratic
indices of the adjoint and symmetric representation of $O(N) \subset U(N)$,
normalized so that $I_2({\tt fund.}) = 1$.

To have a gauge-invariant theory we must add a term to the classical
action which is not gauge invariant, but whose gauge variation will
cancel these anomalies due to the fermions\footnote{We thank Massimo
Porrati for an invaluable discussion of this point.}.  Consider adding a
Chern-Simons term to the action:
\[
S = \int d^3x \, \nu(y) \epsilon^{\alpha \beta \gamma} {\rm Tr} \bigl(
A_\alpha \partial_\beta A_\gamma + i {2 \over 3} A_\alpha A_\beta
A_\gamma\bigr) \,\,\, .
\]
The Chern-Simons coupling $\nu(y)$ is taken to be piecewise constant
between the 8-branes, but with a discontinuity at the location of
every 8-brane:
\be
\nu(y) = \nu_0 \left( - 8 + \sum_{I=1}^{16} \theta(y - y_I)\right)
\,\,\, .
\label{cscoeff}
\ee
Under a gauge transformation the Chern-Simons
action has an anomalous variation
\[
\delta_\Lambda S = \int \nu(y) \, d \, {\rm Tr} (\Lambda dA)\,.
\]
Integration by parts turns this into a sum of localized contributions:
\[
\delta_\Lambda S = \int {\rm Tr}
\left( 8 \nu_0 (\Lambda dA)\vert_{y=0}
+ 8 \nu_0 (\Lambda dA)\vert_{y=\pi \tRii}
- \nu_0 \sum_{I=1}^{16} (\Lambda dA) \vert_{y = y_I}\right)
\,\,\, .
\]
So including the Chern-Simons term will cancel the fermion anomalies,
provided we choose the coefficient $\nu_0 = -1/4\pi$.

The presence of the Chern-Simons term follows purely from anomaly
considerations in 2+1 dimensions, but it also has a clear spacetime
origin and interpretation, which we now outline.  From the spacetime point of
view the D8-branes are sources for the R-R 10-form field strength,
which plays the role of a cosmological constant in massive
\IIA~supergravity \cite{Romans,Polchinski}.  The 10-form field
strength is piecewise constant between the 8-branes, but jumps at the
location of an 8-brane \cite{polchinskiwitten}.  This is exactly the
behavior of the function $\nu(y)$, which suggests that we should
identify $\nu(y)$ with the (dual of) the R-R 10-form field strength.

This identification can be made precise by noting that the R-R 10-form
indeed couples to a 2-brane via a Chern-Simons term
\cite{GreenHullTownsend}.  This coupling can be heuristically
motivated by expanding the coupling between the R-R $(p+1)$-form
potentials $C^{(p+1)}$ and the 2-brane world-volume field strength $F$
\cite{Li,DouglasCouplings,Iflow}, and integrating by parts:
\beas
S & = & \int C \, {\rm Tr} \, e^{F} \\
  & = & \int C^{(3)} + C^{(1)} {\rm Tr} (F) + \half C^{(-1)} {\rm Tr}
         \left(F^2\right) \\
  & = & \int \cdots + \half d C^{(-1)} {\rm Tr} \left(AdA + i {2 \over 3} A^3
         \right)
\eeas
We can formally regard $dC^{(-1)}$ as the dual of the 10-form field
strength, and identify it with $\nu(y)$.  This provides the spacetime
derivation of the Chern-Simons coupling we found above from
world-volume gauge theory considerations.

\subsection{Supersymmetry Anomaly Cancellation}

The next step is to build a supersymmetric action which incorporates
the Chern-Simons term.  The Chern-Simons term by itself breaks
supersymmetry, and to restore it we will have to modify the rest of
the M(atrix) theory action in an appropriate way.  The spacetime
interpretation of this procedure is clear: turning on the Chern-Simons
term is equivalent to turning on a R-R background, which breaks
supersymmetry unless an appropriate NS-NS background is also turned
on.  The reader who does not wish to see the details can skip ahead to
the next sub-section, where we give the results in components.

Let us first outline what we expect to find.  In the infinite momentum
frame, the 32 supersymmetries of M-theory are organized into 16
dynamical supersymmetries that act linearly on the fields, and 16
kinematical supersymmetries that are non-linearly realized
\cite{banksseibergshenker}.  The heterotic compactification on
$({\bf S}^1/\integer_2)\times{\bf S}^1$
breaks half the supersymmetry, so we expect to
find 8 dynamical and 8 kinematical supersymmetries in the M(atrix)
theory.

In this section we will be primarily concerned with achieving
dynamical supersymmetry.  In 2+1 dimensions, in terms of eight spinor
supercharges $\hbox{\cmss Q}_{A=1,\ldots,8}$, the dynamical
supersymmetries are generated by $\bar{\epsilon}_A \hbox{\cmss
Q}_A$, where the $\epsilon_A$ are eight spinor parameters invariant
under the orbifold $\integer_2$ action, $i \gamma^2 \epsilon_A =
\epsilon_A$.  We will refer to this amount of supersymmetry as ${\cal
N} = (0,8)$ in 2+1 dimensions.

Note that ${\cal N}=(0,8)$ supersymmetry does not require
(2+1)-dimensional
Poincar\'e invariance: the supersymmetry algebra closes on
translations in the $t$ and $x$ directions, but does not include
translations in the orbifold direction $y$.  We do, however, expect
the system to have a $Spin(7)$ R-symmetry, corresponding to spatial
rotations in the directions orthogonal to the 2-brane.

We will work in terms of ${\cal N} = 1$ superfields in 2+1 dimensions
(see Appendix A).  This will only allow us to make a $G_2$ subgroup of
$Spin(7)_R$ manifest.  We introduce a vector multiplet
\[
\Gamma_a = i A_\alpha \left(\gamma^\alpha \theta\right)_a + \tbt
\lambda_a
\]
and seven adjoint scalar multiplets
\[
\Phi_i = X_i + \bar{\theta} \psi_i + \half \tbt F_i
\]
where $i=1,\ldots,7$ is an index in the ${\bf 7}$ of $G_2$.  Also
$a=1,2$ is an $SO(2,1)$ spinor index, while $\alpha = 0,1,2$ is an
$SO(2,1)$ vector index.  The hallmark of ${\cal N} = (0,8)$
supersymmetry will be the appearance of an enhanced $Spin(7)_R$
symmetry, under which the seven scalars $X_i$ transform as a vector
and the eight fermions $\lambda,\psi_i$ transform as a
spinor\footnote{The relevant group theory is that the spinor of
$Spin(7)$ decomposes into ${\bf 7} \oplus {\bf 1}$ of $G_2$.
Thus, to construct a
theory with 16 supercharges in terms of ${\cal N}=1$ superfields,
$G_2 \subset Spin(7)_R$ plays the same
role in 2+1 dimensions that $SU(3) \subset Spin(6)_R$ does in 3+1
dimensions.}.

Our starting point is the fact that the usual (Poincar\'e invariant)
${\cal N}=8$ Yang-Mills Lagrangian is the $\tbt$ component of
\[
{\cal F} = {\rm Tr} \, \left(\half W^a W_a - 2 \nabla^a \Phi_i
\nabla_a \Phi_i + f_{ijk} \Phi_i [\Phi_j,\Phi_k] \right)
\]
where $f_{ijk}$ is a suitably normalized totally antisymmetric
$G_2$-invariant tensor.  This leads us to expect a term in the action
\[
S = - {1 \over \gymb} \int d^3x \, z(y) {\cal F}\vert_{\tbt}\,.
\]
Since we do not expect to have translation invariance along the axis
of the cylinder, we have allowed for the possibility that the coupling
constant depends on $y$ through some unknown function $z(y)$.  In the
central region where the cosmological constant $\nu(y)$ vanishes, we
will choose to normalize $z(y) = 1$.

As it stands this action is not supersymmetric.  Under a supersymmetry
variation the $\tbt$ component of ${\cal F}$ changes by a total
derivative, which can then act on $z(y)$.  There is a simple way to
modify any action of this form to restore at least ${\cal N} = (0,1)$
supersymmetry.  Consider the component expansion of ${\cal F}$:
\[
{\cal F} = A + \bar{\theta} \xi + \tbt L \,\,\, .
\]
Under supersymmetry the top component of ${\cal F}$ changes by a total
divergence, $\delta_\epsilon L = - {i \over 2} \bar{\epsilon}\gamma^\alpha
\partial_\alpha \xi$.  So the action built just from $L$ has a
supersymmetry variation (after integrating by parts)
\beas
\delta_\epsilon S & = & - {1 \over \gymb} \int d^3x \, {dz \over dy}
{i \over 2} \bar{\epsilon} \gamma^2 \xi \\
& = & - {1 \over \gymb} \int d^3x \, {dz \over dy}
\left(-\half\right) \bar{\epsilon} \xi \,\,\, .
\eeas
In the second line we used the fact that the spinor parameter
$\epsilon$ satisfies the orbifold projection condition $i \gamma^2
\epsilon = \epsilon$.  But now, noting that the supersymmetry
variation of the bottom component of ${\cal F}$ is $\delta_\epsilon A
= \bar{\epsilon} \xi$, we see that we can compensate for the
supersymmetry variation of this action by adding a term to the
Lagrangian proportional to the bottom component of ${\cal F}$.  This
leads us to the ${\cal N} = (0,1)$ supersymmetric Yang-Mills action
\[
S_{YM} = - {1 \over \gymb} \int d^3x \, \left(z(y) {\cal F}\vert_{\tbt}
+ \half {dz \over dy} {\cal F}\vert_{\theta=0}\right)\,.
\]
This will be one term in our M(atrix) gauge theory action.

Unfortunately, adding the bottom component of ${\cal F}$ to the
Lagrangian breaks the $Spin(7)$ R-symmetry we were hoping to achieve.
To see this, note that the bottom component of ${\cal F}$ has a term
involving the tensor $f_{ijk}$.
\[
{\cal F}\vert_{\theta=0} = {\rm Tr} \left(2 \overline{\lambda}
\lambda - 2 \overline{\psi}_i \psi_i + f_{ijk} X_i [X_j,X_k] \right)
\]
$f_{ijk}$ is invariant under $G_2$, but not invariant under $Spin(7)_R$,
so this term must be canceled.

This can be done by introducing a superpotential.  From ${\cal W} =
{\rm Tr} \, \left(\Phi_i \Phi_i\right)$ we build a $(0,1)$
supersymmetric action:
\[
S_W = - {1 \over \gymb} \int d^3x \, \left(- {4 \over 3} {dz \over dy}
{\cal W}\vert_{\tbt} - {2 \over 3} {d^2z \over dy^2} {\cal W}\vert_{\theta=0}
\right) \,\,\, .
\]
The sum $S_{YM} + S_W$ has a $Spin(7)_R$-invariant potential for the
bosonic fields.  To make this manifest one must eliminate the
auxiliary fields using their equations of motion.
\[
F_i = {3 \over 8} f_{ijk} [X_j,X_k] - {1 \over 3z} {dz \over dy} X_i
\]
The cross term from $\vert F_i \vert^2$ cancels against the unwanted
term $f_{ijk} X_i [X_j,X_k]$ which appeared in ${\cal F}\vert_{\theta=0}$.

We have not yet achieved a $Spin(7)_R$ symmetry which includes the
fermions.  To see this the relevant terms in the action are
\bee
S_{YM} + S_W & = & - {1 \over \gymb} \int d^3x \, {\rm Tr}
\biggl\lbrace
z(y) \left(
-2i \overline{\lambda} \gamma^\alpha D_\alpha \lambda
-2i \overline{\psi}_i \gamma^\alpha D_\alpha \psi_i \right) \nonumber \\
& & \qquad \qquad \qquad \qquad + {dz \over dy} \,
\left(\overline{\lambda}
\lambda - {1 \over 3} \overline{\psi}_i \psi_i \right) + \cdots
\biggr\rbrace
\label{relevantterms}
\eee
which arise from the fermion kinetic terms, the bottom component of
${\cal F}$, and the top component of ${\cal W}$.  The fermion kinetic
terms have a $Spin(7)_R$ R-symmetry, but this does not extend to the
rest of the action.

This can be fixed by introducing a Chern-Simons term.  The
Chern-Simons term appears in the top component of a superfield ${\cal
G}$ which
is given in Appendix A.  We introduce the action
\beas
S_{CS} & = & \int d^3x \, \nu(y) {\cal G}\vert_{\tbt} \\
& = & \int d^3x \, \nu(y) {\rm Tr} \left\lbrace
\epsilon^{\alpha\beta\gamma} \, \left(
A_\alpha \partial_\beta A_\gamma + i {2 \over 3} A_\alpha A_\beta A_\gamma
\right) + \overline{\lambda}
\lambda \right\rbrace \,\,\, .
\eeas
The Chern-Simons coupling $\nu(y)$ was determined in the previous
section from gauge anomaly cancellation.
\[
\nu(y) = - {1 \over 4 \pi} \left( - 8 + \sum_{I=1}^{16} \theta(y - y_I)
\right) \,\,\, .
\]
If we add the Chern-Simons action to the action (\ref{relevantterms}), the
result will be a $Spin(7)_R$-invariant potential for the fermions,
provided that
\be
\nu(y) =  {4 \over 3 \gymb} {dz \over dy} \,\,\, .
\label{cscouplings}
\ee
With this choice, the action $S_{YM} + S_W + S_{CS}$ has a $Spin(7)$
R-symmetry.  We will be able to make this manifest in the next section
when we write the action out in components.

But the Chern-Simons action $S_{CS}$ is not supersymmetric, because the
top component of the superfield ${\cal G}$ changes by a total derivative
under supersymmetry transformation
\[
\delta_\epsilon S_{CS} = \int d^3x \, \nu(y)
\partial_\alpha \left[ - i \epsilon^{\alpha\beta\gamma} {\rm Tr}
\, \left(\bar{\epsilon} \gamma_\beta \lambda A_\gamma\right)\right]
\,\,\, .
\]
Integration by parts turns this into a sum of contributions localized
at the positions of the D8-branes, plus surface terms at the ends of
the cylinder.  For spinors satisfying $i \gamma^2 \epsilon = \epsilon$
the result can be expressed as
\[
\delta_\epsilon S_{CS} = \int d^2x \, \left(8 \, {1 \over 8 \pi}
{\rm Tr} \left(A_-\delta_\epsilon A_+\right)\vert_{y=0,\pi\scriptRii}
- {1 \over 8 \pi} \sum_{I=1}^{16} {\rm Tr} \left(A_-\delta_\epsilon A_+
\right)\vert_{y=y_I}\right) \,\,\, .
\]
The trick we used before to restore supersymmetry, namely adding the
bottom component of the superfield ${\cal G}$ to the action, does not
work for the Chern-Simons term --- the bottom component is not gauge
invariant (indeed it vanishes in Wess-Zumino gauge).

What saves us are the chiral fermions $\chi_I$ which propagate at the
locations of the D8-branes.  The fields $\chi_I$ are supersymmetry
singlets.  Their classical action (\ref{ChiAction}) only depends on
$A_-$, so naively they only couple to the $A_-$ component of the gauge
field, which is also a singlet under $(0,8)$ supersymmetry.  But the
effective action $\Gamma_I[A_+,A_-]$ obtained by integrating out
$\chi_I$ is {\em not} just a functional of $A_-$.  Instead, it must be defined
to include a local contact term which couples $A_+$ to
$A_-$~\cite{Jackiw}.
This is a consequence of the anomaly (\ref{ChiAnomaly}),
which can be expressed in the equivalent form
\[
D_- {\delta \Gamma_I \over \delta A_-} + D_+ {\delta \Gamma_I \over \delta
A_+} = {1 \over 8 \pi} \left(\partial_- A_+ - \partial_+ A_-\right).
\]
This equation can only be satisfied if $\Gamma_I$ is defined to include
a contact term,
\be
\Gamma_I[A_+,A_-] = {1 \over 8\pi} \int d^2x \, {\rm Tr} \left(
A_+ A_- \right) + \widetilde{\Gamma}_I [A_-]
\label{contactterm}
\ee
where $\widetilde{\Gamma}_I$ is an arbitrary functional
of $A_-$.  From this we see that $\Gamma_I$ has an anomalous variation
under supersymmetry:
\[
\delta_\epsilon \Gamma_I = {1 \over 8 \pi} \int d^2x \, {\rm Tr}
\left(A_- \delta_\epsilon A_+\right)\vert_{y = y_I}
\,\,\, .
\]
This is precisely what is needed to cancel the supersymmetry variation
of the Chern-Simons term at the location of the D8-branes.  A similar
supersymmetry anomaly arises from the $y$-independent modes of the
fields $\psi_A$, as required by their gauge anomaly
(\ref{PsiAnomaly}).  This will cancel against the surface terms
which appeared at the boundaries of the \tIA~orbifold in the
supersymmetry variation of the Chern-Simons term.

The action we have determined, $S_{YM} + S_W + S_{CS} + S_\chi$, is
both supersymmetric and $Spin(7)_R$-invariant.  It must, therefore,
have the desired ${\cal N} = (0,8)$ supersymmetry.  In the next
section we write out the full action and its $(0,8)$ supersymmetry
transformations in components.

\subsection{Component Expansion of the M(atrix) Action}

In the previous sections we constructed a gauge invariant ${\cal N} =
(0,8)$ supersymmetric action.  We now collect our results, and expand
the action in components.  This will allow us to write it in a way
that makes the $Spin(7)_R$ R-symmetry manifest.

In obtaining the action we were led to introduce two functions
\beas
\nu(y) & = & - {1 \over 4 \pi} \left( -8 + \sum_{I=1}^{16}
\theta(y-y_I)\right) \\
\noalign{\vskip 0.2 cm}
z(y) & = & {\rm const.} - {3 \gymb \over 32 \pi} \sum_{I=1}^{16}
\left\vert y - y_I \right\vert,
\eeas
where the $y_I$ are the positions of the D8-branes in type \tIA.  The
constant is to be chosen so that $z(y) = 1$ in the central region
where $\nu(y) = 0$.

We emphasize that we have determined these functions by demanding
supersymmetry and gauge anomaly cancellation in 2+1 dimensions.  But
they do have a clear spacetime interpretation: they correspond to the
background fields established by the D8-branes.  As one might expect,
$z(y)$ gives the spatial variation of the NS-NS fields, just as
$\nu(y)$ was shown to give the variation of the R-R 10-form in section
3.1.  The precise identification comes from noting
that in a system of parallel D8-branes the background fields vary
according to \cite{polchinskiwitten}
\[
\begin{array}{ll}
\hbox{\rm dilaton} & \enskip e^\phi \sim z^{-5/6}(y) \\
\noalign{\vskip 0.3 cm}
\hbox{\rm metric} & \enskip G_{\mu \nu} \sim z^{-1/3}(y) \, \eta_{\mu\nu} \\
\noalign{\vskip 0.3 cm}
\hbox{\rm R-R 10-form} & \enskip F^{(10)} \sim \nu(y) \, dx^0 \cdots dx^9.
\end{array}
\]

Under $Spin(7)_R$, the seven scalars $X_i$ transform as a vector,
while the eight fermions $\lambda$, $\psi_i$ transform as a spinor.
At this point it is useful to make a field redefinition.  We define
\beas
Y^i & = & z^{1/3}(y) X_i \\
\Psi_A & = & \left\lbrace z^{1/3}(y) \lambda,\,\,
z^{1/3}(y) \psi_i\right\rbrace \,\,.
\eeas
>From the M(atrix) gauge theory point of view these redefinitions can
be motivated by noting that certain properties of the action -- the
existence of flat directions, as well as kinematical supersymmetry --
look most natural in terms of the rescaled variables\footnote{We are
grateful to E.~Witten for bringing this issue to our attention.}.

These redefinitions can also be given a simple spacetime motivation.
Within the M(atrix) gauge theory the $Spin(7)_R$ vector indices $i,j$
are of course raised and lowered with the flat metric $\delta_{ij}$.
But from the spacetime point of view it is natural to regard the
scalars $X$ as having covariant indices.  This is because in
constructing a supersymmetric Yang-Mills theory these scalars arise
from dimensional reduction of a one-form.  But the embedding
coordinates for a membrane $Y$ should have contravariant indices.  So
they are related by a factor of the inverse metric, $Y^i = z^{1/3}(y)
X_i$.

Collecting our results of the previous two sections, expanding in
components, and integrating out the auxiliary fields, we find the
M(atrix) model action\footnote{Notation: $i,j$ are $Spin(7)_R$ vector
indices, $A,B$ are $Spin(7)_R$ spinor indices, $\alpha,\beta$ are
$SO(2,1)$ Lorentz indices, and $I$ labels the D8-branes.  The metric
is $(-++)$ and $\epsilon^{012} = + 1$.}
\bea
\label{MatrixAction}
S & = & - {1 \over \gymb} \int d^3x \, {\rm Tr} \bigg\lbrace
z(y) F_{\alpha\beta} F^{\alpha\beta} + 2 z^{1/3}(y) D_\alpha Y_i
D^\alpha Y^i - z^{-1/3}(y) [Y_i,Y_j][Y^i,Y^j] \nonumber \\
& & \quad \qquad \qquad \qquad - 2 i z^{1/3}(y)
\overline{\Psi}_A \gamma^\alpha D_\alpha \Psi_A - {d z^{1/3} \over
dy} \overline{\Psi}_A \Psi_A + 2 i \overline{\Psi}_A \gamma^i_{AB}
[Y_i, \Psi_B] \nonumber \\
& & \quad \qquad \qquad \qquad - {4 \over 3} {dz \over dy}
\epsilon^{\alpha \beta \gamma}
\bigl(A_\alpha \partial_\beta A_\gamma + i {2 \over 3} A_\alpha
A_\beta A_\gamma \bigr) \bigg\rbrace \\
& & + i \sum_{I=1}^{16} \int d^3x \delta(y - y_I)
\overline{\chi}_I \left(\partial_- + i A_- + i B_-^{(I)}\right) \chi_I
\,\,\,. \nonumber
\eea
Here $A_\alpha$ is a $U(N)$ gauge field, coupled to adjoint scalars
$Y^i$ and adjoint fermions $\Psi_A$.  The fields $\chi_I$ are the
twisted sector fermions, complex chiral fermions in the fundamental
representation of $U(N)$.  The field $\chi_I$ is also charged under
the $U(1)$ gauge field $B^{(I)}$ which propagates on the $I^{th}$
D8-brane.

This action has several symmetries, all of which are necessary for a
sensible M(atrix) theory interpretation.  As discussed in section 3.1,
the gauge anomalies cancel, and the action enjoys a $U(N)$ gauge
invariance.  Also, as shown in section 3.2, the action is invariant
under the ${\cal N} = (0,8)$ dynamical supersymmetry transformation
\bea
\label{DynamicalSusy}
\delta_\epsilon A_\alpha &=& {i \over 2} z^{-1/3}(y) \bar{\epsilon}_A
\gamma_\alpha \Psi_A \nonumber \\
\delta_\epsilon Y^i &=& - {1 \over 2} \bar{\epsilon}_A \gamma^i_{AB}
\Psi_B \nonumber \\
\delta_\epsilon \Psi_A &=& - {1 \over 4} z^{1/3}(y) F_{\alpha \beta}
\gamma^{\alpha \beta} \epsilon_A - {i \over 2} D_\alpha Y_i \gamma^\alpha
\gamma^i_{AB} \epsilon_B - {i \over 4} z^{-1/3}(y) [Y_i,Y_j]
\gamma^{ij}_{AB} \epsilon_B \\
\delta_\epsilon \chi_I &=& 0 \nonumber
\eea
where the spinor parameter satisfies the $\integer_2$ projection
condition $\epsilon_A = i \gamma^2 \epsilon_A$.  There is an associated
$Spin(7)$ R-symmetry, which corresponds to spatial rotations in the
directions orthogonal to the 2-branes.

The action is also invariant under the ${\cal N} = (0,8)$
kinematical supersymmetry
\be
\label{KinematicalSusy}
\delta_\eta \Psi_A = \eta_A \identity
\ee
where the spinor parameter satisfies $\eta_A = - i \gamma^2 \eta_A$,
and all other variations vanish\footnote{The fact that the
kinematical and dynamical supersymmetry parameters satisfy
opposite-sign $\integer_2$ projections was discussed in section 2.2.}.
Kinematical supersymmetry was discussed in the context of type
\IIA~matrix theory in section 2.1.  It corresponds to those spacetime
supersymmetries which are broken (non-linearly realized) due to the
choice of infinite momentum frame \cite{banksseibergshenker}.  As
such, it should also hold in heterotic M(atrix) theory.  This is
easily verified, but note that there is a non-trivial cancellation
between the variations of the fermion mass and kinetic terms.  Since
kinematical supersymmetry did not play a role in our construction of
this action, it could be regarded as an accidental symmetry, but one
that is necessary to have a sensible M(atrix) theory interpretation.

Finally, the action is invariant under transverse translations
\[
Y_i \rightarrow Y_i + c_i \identity \,\,.
\]
This is simply a spatial translation in the directions orthogonal to
the 2-branes.  A related observation is that the action has the flat
directions required for M(atrix) theory: there are low energy
configurations in which the matrices $Y_i$ are diagonal, corresponding
to widely separated 2-branes.  Again, note that this property did not
play a role in our construction of the action.

\section{BPS States and T-duality}

Having established the structure of heterotic M(atrix) theory, we now
analyze its BPS excitations, and compare them to the heterotic string.
We do this for several reasons.  Showing that the BPS spectra agree
gives evidence that the M(atrix) theory we have described really does
provide a non-perturbative definition of the heterotic string.  Also,
by matching the BPS spectra, we can relate the parameters of M(atrix)
theory to the parameters of the heterotic string.  And finally, this
will show that heterotic T-duality is realized in M(atrix) theory as a
novel form of electric--magnetic S-duality.

The heterotic compactification of M-theory on $\left({\bf S}^1 /
\integer_2\right) \times {\bf S}^1$ gives rise to a richer spectrum of
BPS states than toroidal compactification.  This is due to the
existence of a twisted sector.  Besides the closed membranes of the
untwisted sector, there are membranes with boundaries at the orbifold
fixed points.  These so-called twisted membranes correspond to the
charged excitations present in type \IA~and heterotic string theory.

For future reference, we list the BPS excitations of the heterotic
string, and their corresponding states in M-theory and M(atrix)
theory.  Recall that M-theory is compactified on $\left({\bf S}^1 /
\integer_2\right) \times {\bf S}^1$, where $x = x^1$ is the orbifold
direction and $y = x^2$ is the circle direction (note that these
directions are interchanged in M(atrix) theory).  The $x^{11}$
direction along which M-theory is boosted is compactified on a
regulator circle of radius $R_{11}$.  All the states we discuss are
assumed to have momentum $p_{11} = N / R_{11}$ around the regulator
circle.
\begin{enumerate}
\item {\tt winding mode excitations} \\
A heterotic string wound on ${\bf S}^1$ is realized in M-theory as an
untwisted membrane wrapped on $x^1$-$x^2$.  In the M(atrix) gauge
theory this is realized as a quantum of magnetic flux ${\bf B}$.
\item {\tt momentum mode excitations} \\
A heterotic string with momentum around ${\bf S}^1$ is realized as an
M-theory graviton with momentum around the $x^2$ circle direction.
In M(atrix) theory it corresponds to a quantum of electric flux
${\bf E}_2$ in the $y$ orbifold direction.
\item {\tt charged excitations} \\
Charged heterotic strings correspond to twisted membranes in M-theory.
In M(atrix) theory they arise as excitations of the twisted sector
fermions $\chi_I$.
\item {\tt winding modes on $x^{11}$} \\
The regulator circle is one of the spatial directions in the heterotic
string.  A heterotic string wound on this circle corresponds to a
longitudinal membrane in M-theory, wrapped on $x^1$-$x^{11}$.  In
M(atrix) theory this corresponds to a photon propagating around the
$x$ circle direction.
\end{enumerate}
\noindent
The first three types are the basic BPS states of interest to us.
They can be combined to form additional BPS states, namely charged
winding states and charged Kaluza-Klein states.  The fourth type of
BPS state becomes infinitely heavy as $R_{11} \rightarrow \infty$, but
will be useful later
for normalizing the M(atrix) theory parameters.

\subsection{BPS States for Unbroken $E_8 \times E_8$}

We proceed to analyze the BPS states in M(atrix) theory.  In this
subsection we assume that $E_8 \times E_8$ is unbroken.  Recall that
in M(atrix) theory this corresponds to a configuration in which a
background Wilson line $B_1 = \left(0^8, ({\scriptscriptstyle 1 \over
2})^8\right)$ is turned on and 8 twisted sector fermions $\chi_I$ are
located at each orbifold boundary.  We discuss three topics in turn:
flux quantization, the BPS conditions and equations of motion, and the
spectrum of BPS states.

\subsubsection{Flux Quantization}

In this section we discuss the allowed $U(N)$ bundles in M(atrix)
theory, and the corresponding quantization conditions on electric and
magnetic flux.

We begin by constructing the general $U(N)$ bundle on ${\bf T}^2$.  To
describe this, we introduce the $U(N)$ matrix
\[
V(y) = \left(
\begin{array}{cccccc}
0 & 1 &   &                 &   &   \\
  & 0 & 1 &                 &   &   \\
  &   &   & \ddots          &   &   \\
  &   &   &                 & 0 & 1 \\
e^{- i y / \tRii} & & & &   & 0
\end{array}\right) \,\,\, .
\]
As $y$ ranges from $0$ to $2 \pi \tRii$, $V(y)$ traces out a
non-contractable loop in $U(N)$, which generates $\pi_1(U(N)) =
\integer$.  The possible $U(N)$ bundles on ${\bf T}^2$ are labeled by
an integer $m$, and can be described through the introduction of
twisted boundary conditions.  For a field $\phi(x,y)$ in the
fundamental representation of $U(N)$, the boundary conditions are
\beas
\phi(x+2\pi\tRi,y) &=& V^m(y) \phi(x,y) \\
\phi(x,y+2\pi\tRii) &=& \phi(x,y) \,\,\, .
\eeas
These conditions are gauge-equivalent to 't Hooft's boundary
conditions for $m$ torons \cite{tHooft}, but have been chosen to be
compatible with the $\integer_2$ action of $\Omega \cdot P$, given in
(\ref{OmegaAction}), (\ref{PAction}).  Then we can make a $\integer_2$
projection by $\Omega \cdot P$ to obtain a $U(N)$ bundle on ${\bf S}^1
\times \left({\bf S}^1 / \integer_2\right)$.  This results in boundary
conditions on the M(atrix) theory fields
\beas
A_\alpha(x + 2 \pi \tRi, y) & = & V^m(y) \left( A_\alpha (x, y)
- i \partial_\alpha \right) V^{-m}(y) \\
Y^i(x + 2 \pi \tRi, y) & = & V^m(y) Y^i (x, y) V^{-m}(y) \\
\chi_I(x + 2 \pi \tRi) & = & V^m(y_I) \chi_I(x)
\eeas
besides those given in (\ref{bc}).  It follows from these boundary
conditions that magnetic flux is quantized, with the first Chern class
taking on half-integer values due to the $\integer_2$ orbifold:\footnote{
We use the notation ${\bf B} = F_{12}$, ${\bf E}_1 = F_{01}$,
${\bf E}_2 = F_{02}$.}
\be
\label{Bflux}
{1 \over 2 \pi} \int_{{\widetilde C}_2} d^2x \, {\rm Tr} {\bf B}
= {m \over 2} \,\,\, .
\ee

A minimum-action classical field configuration which respects these
boundary conditions is
\bea
\label{ToronConfig}
A_1(x,y) &=& 0 \nonumber \\
A_2(x,y) &=& {m x \over 2 \pi N \tRi \tRii} \identity + {1 \over N \tRii}
{\rm diag.} (1,2,\ldots,N) \,\,\, .
\eea
Note that the energy of this field configuration comes entirely from the
center of mass $U(1)$ part of the field.

Large gauge transformations, generated by $V(y)$, imply that $A_2$ is
a periodic variable: $A_2 \approx A_2 + {1 \over N \tRii} \identity$.
This can be easily seen by shifting $x \rightarrow x + 2 \pi \tRi$ in
(\ref{ToronConfig}).  This periodicity leads to a quantization
condition on the conjugate momentum.
\be
\label{Eflux}
\int_0^{\pi \tRii} dy \, {\rm Tr} {\bf E}_2 = {\gymb \tRii \over 4 \tRi}
\, n\,, \hspace{0.8 cm} n \in \integer \,\,\, .
\ee
Again note that a minimum-energy field configuration with electric
flux gets all its energy from the center of mass $U(1)$ part of the
field.  The generalization of this quantization condition when a
Wilson line is turned on will be discussed in section 4.2.

\subsubsection{BPS Conditions and Equations of Motion}

In this subsection we give the BPS conditions and classical equations
of motion for M(atrix) theory.  As pointed out in the previous
subsection, minimum-energy field configurations with electric and
magnetic flux get all their energy from the $U(1)$ part of the fields.
So we only need to consider the BPS conditions and equations of motion
for the $U(1)$ part of the action (\ref{MatrixAction}).

A BPS state should be invariant under a linear combination of the
${\cal N} = (0,8)$ kinematical and dynamical supersymmetries.  So we
require the supersymmetry variation of the fermions (\ref{DynamicalSusy}),
(\ref{KinematicalSusy})
\be
\delta_\epsilon \Psi_A + \delta_\eta \Psi_A = {1 \over 2} z^{1/3}(y)
\left(\begin{array}{cc}
- {\bf E}_1 & {\bf E}_2 - {\bf B} \\
{\bf E}_2 + {\bf B} & {\bf E}_1
\end{array} \right)
\left(\begin{array}{c} 0 \\ \epsilon_{2A} \end{array} \right)
+ \left(\begin{array}{c} \eta_{1A} \\ 0 \end{array} \right)
- {i \over 2} \partial_\alpha Y_i \gamma^\alpha \gamma^i_{AB}
\epsilon_B
\label{susycond}
\ee
to vanish for non-trivial spinor parameters $\epsilon_A$, $\eta_A$.
In writing out the supersymmetry transformation we have assumed that
the spinors satisfy the $\integer_2$ projection conditions $\epsilon_A
= i \gamma^2 \epsilon_A$, $\eta_A = -i \gamma^2 \eta_A$ discussed in
section 2.2.  Setting $z(y) = 1$ to correspond to unbroken $E_8 \times
E_8$, this leads to the BPS conditions
\bee
{\bf E}_1 & = & 0 \nonumber \\
{\bf E}_2 - {\bf B} & = & {\rm const.} \nonumber \\
Y_i & = & {\rm const.}
\label{bpscondition}
\eee
which fortunately are compatible with the boundary conditions
(\ref{bc}).

A curious feature of $(0,8)$ supersymmetry is that states which
satisfy the BPS conditions are not necessarily solutions to the
equations of motion -- note that ${\bf E}_2+{\bf B}$ is left
undetermined by the BPS conditions.  This is not surprising, since the
$(0,8)$ supersymmetry algebra does not include translations in the
orbifold $y$ direction.  In order to completely determine the fields
we must look at the equations of motion and the Bianchi identity.  In
components, the equations of motion are\footnote{A source term on the
right hand side of these equations, discussed in section 4.2, will be
important when a Wilson line is turned on.}
\beas
\partial_x {\bf E}_1 + \partial_y {\bf E}_2 & = & 0 \\
\partial_t {\bf E}_1 + \partial_y {\bf B} & = & 0 \\
\partial_t {\bf E}_2 - \partial_x {\bf B} & = & 0
\eeas
while the Bianchi identity is
\[
\partial_x {\bf E}_2 - \partial_y {\bf E}_1 - \partial_t {\bf B} = 0 \, .
\]
So the the only static field configurations consistent with the BPS
conditions are ${\bf E}_1 = 0$, ${\bf E}_2 = {\rm const.}$, ${\bf B}
= {\rm const.}$

\subsubsection{BPS Spectrum}

We are now in a position to identify the BPS states in M(atrix) theory
for unbroken $E_8 \times E_8$.

First we discuss the origin of gauge quantum numbers.  This problem
has been studied in a T-dual version by Kachru and Silverstein
\cite{kachrusilverstein}; see also \cite{lowe, sjrey}.
Based on these works,
we expect to find that when $N$ is odd the M(atrix) theory will have
BPS states in the $({\bf 128},{\bf 1})$ and $({\bf 1}, {\bf 128})$ of
$SO(16) \times SO(16)$, localized near $A_1 = 0$ and $A_1 = 1 / 2
\tRi$, respectively.  When $N$ is even, we expect to find
gauge-neutral BPS states with wavefunctions that are smeared out in
$A_1$, as well as charged BPS states, localized near $A_1 = 0$ and $1
/ 2 \tRi$, in the $({\bf 120, 1})$ and $({\bf 1, 120})$.

Gauge quantum numbers are generated by the twisted sector fermions $\chi$.
For unbroken $E_8 \times E_8$ their action is (see ~(\ref{twisted}))
\beas
S_{\rm twisted} & = & i \int d^2x \Big\lbrace \,\,
\chi^{(1)} \left(\partial_- - i \left.A_1\right\vert_{y = 0}
- i B^{(1)}_1 \right) \chi^{(1)} \\
& & \qquad\qquad\!\!\! + \chi^{(2)} \left(
\partial_- - i \left.A_1\right\vert_{y = \pi \tRii}
- i B^{(2)}_1 \right) \chi^{(2)}\,\, \Big\rbrace
\eeas
where $\chi^{(1)}$ and $\chi^{(2)}$ are the twisted sector fermions
located at $y = 0$ and $y = \pi \tRii$, respectively.
$\left(B^{(1)}_1, B^{(2)}_1\right) = \left(0^8,( 1
/ 2 \tRi)^8\right)$ are the background $SO(16)$ gauge fields at $y =
0$ and $y = \pi \tRii$, and we have set $A_0 = B_0 = 0$.

Let's see how this works for $N=1$.  Then the M(atrix) gauge group is
$U(1)$, broken to $O(1) \subset U(1)$ on the boundaries.  In a BPS
configuration, $A_1$ should be constant.  But then the boundary
conditions (\ref{bc}) force $A_1$ to lie in an $O(1)$ subgroup of
$U(1)$.  There are only two possibilities: either $A_1 = 0$ or $A_1 =
1 / 2 \tRi$.  This is the T-dual version of the observation that in
type \IA, a single unpaired D0-brane must be locked at one of the two
orientifolds \cite{kachrusilverstein}.

If $A_1 = 0$, then the 16 real fermions $\chi^{(1)}$ have zero modes,
and generate a $({\bf 128},{\bf 1})$ of $SO(16) \times SO(16)$.  On
the other hand, if $A_1 = 1/2\tRi$, then $\chi^{(2)}$ has zero modes,
and we get a $({\bf 1},{\bf 128})$.  Note that we only get 128 states
because the discrete $O(1)$ gauge symmetry $\chi \rightarrow - \chi$
kills states containing an odd number of $\chi$ excitations.

When $N=2$ the M(atrix) gauge group is $U(2)$, broken to $O(2)$ on the
boundaries.  The BPS conditions do not fix $A_1$ discretely, but
rather allow it to be an arbitrary element of $O(2)$.  For generic
values of $A_1$, there are no $\chi$ zero modes, and we expect the
wavefunction to spread out in $A_1$ in order to satisfy the BPS
condition ${\bf E}_1 = 0$.  Such a configuration corresponds to a
gauge-neutral state of the heterotic string.  But when $A_1 = 0$ there
are 16 real doublets of fermion zero modes in $\chi^{(1)}$.  Denoting
these fields $\chi_M^{(1)} = \left(\begin{array}{c}\chi^1_M \\
\chi^2_M\end{array}\right)$, where $M=1,\ldots,16$ is an
$SO(16)$ index, we can make states
\[
\left(\chi^1_M + i \chi^2_M \right)
\left(\chi^1_N + i \chi^2_N \right)\vert 0 >
\]
in the $({\bf 120},{\bf 1})$ of $SO(16) \times SO(16)$.  These states
are localized near $A_1 = 0$.  Likewise, when $A_1 = 1/2\tRi$, we can
make states in the $({\bf 1},{\bf 120})$ from the zero modes in
$\chi^{(2)}$.  States with additional (even) numbers of $\chi$
excitations are expected to be unstable.

This pattern of gauge quantum numbers is expected to repeat in $N$
modulo 2~\cite{kachrusilverstein}, and hence $E_8$ multiplets should
be present in the large-$N$ limit.  Alternatively, one may give an
interpretation to the results for finite $N$ by compactifying a null
direction~\cite{DLCQ}.

Next we discuss the spectrum of heterotic momentum and winding states.
These correspond to states in M(atrix) theory with electric and
magnetic flux, superimposed on top of any gauge quantum numbers that
may be present.  As mentioned previously, the energy of these states
comes purely from the center of mass $U(1)$ part of the fields.  So all
we need is the $U(1)$ Hamiltonian
\[
H_{U(1)} = {2 \over \gymb} \int_{{\widetilde C}_2} d^2x \,
\big({\bf E}_1^2 + {\bf E}_2^2 + {\bf B}^2\big)\,.
\]
We use the quantization conditions (\ref{Bflux}), (\ref{Eflux}) for
the gauge group $U(1)$.  This should give the correct spectrum of
fluxes for any $U(N)$ M(atrix) gauge group, provided that we put an
overall factor of $1/N$ in front of the Hamiltonian.  So a BPS state
with $n$ units of electric flux and $m$ units of magnetic flux has an
energy
\be
\label{fluxes}
E_{nm} = {1 \over 2 N \tRi} \left({\gymb \tRii \over 2} \, n^2
+ {2 \over {\gymb \tRii} } \, m^2 \right)\,.
\ee

Using the dictionary (\ref{HetParams}), (\ref{tIAParams}),
(\ref{YMcoupling}) between the heterotic and M(atrix) parameters, we
can re-express this energy in terms of heterotic variables.
\[
E_{nm} = {R_{11} \over 2 N} \left(\left(n \over R_H\right)^2 +
\left(m R_H \over \alpha'_H\right)^2\right)
\,\,\, .
\]
This is the correct light-front energy of a heterotic string state
with $N$ units of longitudinal momentum, provided that we identify $n$
with the momentum quantum number of the heterotic string, and $m$ with
the winding quantum number.  This state satisfies heterotic
level-matching \cite{DVV, sjrey}~only if either $n$ or $m$ vanishes.

The heterotic spectrum is invariant under T-duality, which exchanges
momentum and winding and inverts the radius of the circle:
\beas
n & \leftrightarrow & m \\
R_H & \leftrightarrow & \alpha'_H / R_H
\,\,\, .
\eeas
Perforce this is also a symmetry of the M(atrix) theory spectrum.  But
note that in M(atrix) theory terms it is a symmetry which exchanges
electric and magnetic flux, and simultaneously inverts the
dimensionless coupling constant:
\beas
n & \leftrightarrow & m \\
\noalign{\vskip 0.1cm}
{\gymb \tRii \over 2} & \leftrightarrow & {2 \over \gymb \tRii}
\,\,\, .
\eeas
This is electric--magnetic S-duality in 2+1 dimensions.  Normally
electric--magnetic duality is not possible in (2+1) dimensions,
because the number of independent components of the electric and
magnetic fields are not the same.  It is only made possible in
M(atrix) theory by the $\integer_2$ orbifold projection, which
eliminates states having electric flux ${\bf E}_1$.

Aspinwall~\cite{aspinwall} and Schwarz~\cite{schwarz} have shown that
M-theory compactified on a shrinking ${\bf T}^2$ gives rise to
ten-dimensional type \IIB~string theory, by opening up an `extra'
dimension corresponding to membrane wrapping modes.  In M(atrix)
theory this is a consequence of S-duality in (3+1)-dimensional ${\cal
N} = 4$ gauge theory~\cite{sethisusskind}, or alternatively is due to
the existence of a nontrivial superconformal fixed point with
$Spin(8)_R$ R-symmetry~\cite{banksseiberg}.  M-theory compactified on
$\left({\bf S}^1/\integer_2 \right) \times {\bf S}^1$ exhibits similar
behavior.  As the volume of the orbifold shrinks, membrane wrapping
states become continuous~\cite{horavawitten}, and give rise to an
extra dimension.  We see here that this is realized in M(atrix) theory
via (2+1)-dimensional S-duality. Because of this, we expect that the
enhancement of R-symmetry to $Spin(8)_R$ persists even when Wilson lines
are turned on.

\subsection{BPS States: Turning on Wilson Lines}

The BPS spectrum of the M(atrix) theory should be modified when a
heterotic Wilson line is turned on.  This comes about as a result of
some rather intricate dynamics.  Recall that turning on a heterotic
Wilson line modifies the M(atrix) gauge theory, by moving the twisted
sector fermions $\chi_I$ away from the orbifold boundaries, and by
inducing position-dependent Yang-Mills and Chern-Simons couplings in
the rims of the orbifold, that is, between the positions of the
fermions $y_I$ and the orbifold boundaries.

This produces two important effects, which we analyze in this section.
\begin{itemize}
\item
The position-dependent Yang-Mills and Chern-Simons couplings modify
the behavior of electric and magnetic fields in the rims of the
orbifold.  This is responsible for Wilson line deformation of the
spectrum of untwisted states in M-theory (corresponding to uncharged
states of the heterotic string).
\item
The $\chi_I$ zero modes are charged under the M(atrix) theory gauge
group.  Exciting these zero modes produces electric and magnetic
fields in the rims of the orbifold, at a cost in energy when the
fields $\chi_I$ are moved away from the boundaries.  This is
responsible for Wilson line deformation of the spectrum of twisted
states in M-theory (seen as gauge symmetry breaking in the heterotic
string).
\end{itemize}

We first note that the quantization conditions on electric and
magnetic flux are slightly modified when the Wilson line is turned on.
A state with $m$ units of magnetic flux and $n$ units of electric flux
is specified by the quantization conditions
\bea
\label{FluxQuant}
& & {1 \over 2\pi}
\int_0^{2 \pi \tRi}dx \int_0^{\pi \tRii} dy \,\, {\rm Tr} {\bf B} =
{m \over 2} \nonumber \\
\noalign{\vskip 0.2 cm}
& & \int_0^{\pi \tRii} dy \, z(y) \,\,
{\rm Tr} {\bf E}_2 = {\gymb \tRii \over 4 \tRi} n \,\,.
\eea
The magnetic flux quantization condition isn't modified, since it's
topological and doesn't depend on the dynamics.  The factor of
$z(y)$ which appears in the electric flux quantization condition
reflects the position-dependent Yang-Mills coupling.  One subtle point
about the electric flux quantization condition should be noted.  The
quantization condition actually applies to the momentum ${\bf \Pi}_2$
which is canonically conjugate to the periodic variable $A_2$.  In the
presence of a Chern-Simons term ${\bf \Pi}_2$ is not the same as the
electric field ${\bf E}_2$.  But in the presence of a Chern-Simons
term a wavefunction picks up a phase under the (large) gauge
transformation $A_2 \rightarrow A_2 + {1 \over N \tRii} \identity$,
which means that the spectrum of ${\bf \Pi}_2$ is no longer integer
quantized.  The two effects exactly compensate each other, and the net
result is that the electric flux quantization condition can be written
as above just as if no Chern-Simons term was present.

In the remainder of this section we only consider the action for the
center-of-mass $U(1)$ part of the gauge group.  As in section 4.1.3,
this should give the correct spectrum of electric and magnetic fluxes
for arbitrary $U(N)$ gauge group, provided we put an overall factor of
$1/N$ in front of the Hamiltonian.

\subsubsection{Untwisted States}

We first discuss states in which the zero modes of the fermions
$\chi_I$ are not excited.  This should give us an indication of the
behavior of untwisted states in M-theory, corresponding to
gauge-neutral states of the heterotic string.

When the heterotic Wilson line is turned on, the BPS conditions that
follow from (\ref{susycond}) are
\bea
{\bf E}_1 & = & 0 \nonumber \\
\noalign{\vskip 0.2 cm}
{\bf E}_2 - {\bf B} & = & {{\rm const.} \over z^{1/3}(y)}
\label{SecondBPS}\\
\noalign{\vskip 0.2 cm}
Y_i & = & {\rm const.} \nonumber
\eea
But, just as in section 4.1.2, a state which satisfies the BPS
conditions is not necessarily a solution to the equations of motion.
In particular the behavior of ${\bf E}_2 + {\bf B}$ is not determined
by the BPS conditions.

So we need the gauge field equations of motion.  It is convenient to
begin by integrating out the twisted sector fermions $\chi_I$, which is
equivalent to putting these fermions in their ground state.  The
effective action for the gauge field is given by a fermion determinant
which can be evaluated in closed form \cite{Schwinger}:
\[
\Gamma_I[A] = \int d^2 x \, {1 \over 8 \pi} \left(\partial_+ A_-
{1 \over \scriptlap} \partial_+ A_- + A_+ A_-\right) \,\,\, .
\]
Here $\partial_\pm = \partial_0 \pm \partial_1$ and $\laplace =
\partial_+ \partial_-$.  The contact term in the determinant has been
chosen to satisfy the anomaly equation, see (\ref{contactterm}).
The equations of motion for the gauge field are non-local, but they
are gauge-invariant because we have cancelled the gauge anomalies:
\bea
\label{AbelianEOM}
& & \partial_x {\bf E}_1 + \partial_y {\bf E}_2 +
{1 \over z} {dz \over dy} {\bf E}_2 + {2 \over 3z} {dz \over dy}
{\bf B} = J^0
\nonumber \\
& & \partial_t {\bf E}_1 + \partial_y {\bf B} + {2 \over 3z}
{dz \over dy} {\bf E}_2 + {1 \over z} {dz \over dy} {\bf B}
= J^1 \nonumber \\
& & \partial_t {\bf E}_2 - \partial_x {\bf B} - {2 \over 3z}
{dz \over dy} {\bf E}_1 = 0 \\
\noalign{\vskip 14 pt}
& & J^0 = J^1 = - {\gymb \over 8 \pi z(y)} \sum_{I=1}^{16} \delta(y - y_I)
{1 \over \partial_-} {\bf E}_1 \,\,\, . \nonumber
\eea
The zero mode should be suppressed in defining the kernel
$1/\partial_-$.  We also have the Bianchi identity
\[
\partial_x {\bf E}_2 - \partial_y {\bf E}_1 - \partial_t {\bf B} = 0\,.
\]

Specializing to static configurations and imposing the BPS condition
${\bf E}_1 = 0$ implies that the fields ${\bf E}_2$, ${\bf B}$ are
constant along the $x$ circle direction (this follows from the
Bianchi identity and the third equation of motion).  The remaining two
equations of motion, which determine the dependence of the fields on
$y$, then simplify to
\beas
\partial_y \left({\bf E}_2 + {\bf B} \right) + {5 \over 3z} {dz \over dy}
\left({\bf E}_2 + {\bf B} \right) & = & 0 \\
\partial_y \left({\bf E}_2 - {\bf B} \right) + {1 \over 3z} {dz \over dy}
\left({\bf E}_2 - {\bf B} \right)& = & 0
\,\,\, .
\eeas

The general solution to these equations of motion
\beas
{\bf E}_2(y) & = & {{\bf E}_0 + {\bf B}_0 \over 2 z^{5/3}(y)}
+ {{\bf E}_0 - {\bf B}_0 \over 2 z^{1/3}(y)} \\
\noalign{\vskip 6 pt}
{\bf B}(y) & = & {{\bf E}_0 + {\bf B}_0 \over 2 z^{5/3}(y)}
- {{\bf E}_0 - {\bf B}_0 \over 2 z^{1/3}(y)}
\eeas
saturates the BPS conditions (\ref{SecondBPS}).  ${\bf E}_0$ and ${\bf
B}_0$ are integration constants (the values of the electric and
magnetic fields in the central region where the cosmological constant
vanishes and $z(y) = 1$).  The energy of this state is
\[
H_{U(1)} = {1 \over \gymb} \int_{\widetilde{C}_2} d^2x \, {1 \over z(y)} \,
\left({({\bf E}_0 + {\bf B}_0)^2 \over z^{4/3}(y)} +
{({\bf E}_0 - {\bf B}_0)^2 \over z^{-4/3}(y)} \right)\,.
\]

The constants ${\bf E}_0$ and ${\bf B}_0$ are determined by imposing
the flux quantization conditions (\ref{FluxQuant}).  In terms of the
integral
\[
I_k \equiv \int_0^{\pi \tRii} dy \, z^k(y)
\]
we find that a state with $n$ units of electric flux has an energy
\[
E_n = n^2 \, {\pi \gymb \tRii^2 \over 2 \tRi} \,
{(I_{-1/3})^2 I_{-7/3} + (I_{-5/3})^2 I_{1/3}
\over (I_{-2/3} \, I_{-1/3} + I_{2/3} \, I_{-5/3})^2}
\]
while a state with $m$ units of magnetic flux has an energy
\[
E_m = m^2 \, {2 \pi \over \gymb \tRi} \,
{(I_{2/3})^2 I_{-7/3} + (I_{-2/3})^2 I_{1/3}  \over
(I_{-2/3} \, I_{-1/3} + I_{2/3} \, I_{-5/3})^2 }\,.
\]
As these configurations are BPS saturated, it is natural to expect
that there are no quantum corrections to their classical energies.
The status of such a non-renormalization theorem in (2+1)-dimensional
${\cal N} = (0,8)$ supersymmetry deserves further study, however.  One
would like to be certain that the renormalizations observed
in~\cite{kimrey, banksseibergsilverstein} in ${\cal N} = 8$
supersymmetric quantum mechanics do not take place in 2+1 dimensions.

\subsubsection{Twisted States}

It is also possible to understand, at least in a qualitative way, how
the energy of a twisted state in M-theory is modified when a Wilson
line is turned on.  In the heterotic string, this corresponds to the
fact that the Wilson line breaks the gauge symmetry and gives a mass
to charged states.  In M(atrix) theory it must mean that exciting a
zero mode of $\chi_I$ costs energy.

To see that this is indeed the case, we write the gauge field equations
of motion for the $U(1)$ part of the action (\ref{MatrixAction}).
\beas
& & \partial_x {\bf E}_1 + \partial_y {\bf E}_2 +
{1 \over z} {dz \over dy} {\bf E}_2 + {2 \over 3z}
{dz \over dy} {\bf B} = J^0 \\
& & \partial_t {\bf E}_1 + \partial_y {\bf B} + {2 \over 3z}
{dz \over dy} {\bf E}_2 + {1 \over z} {dz \over dy} {\bf B}
= J^1 \\
& & \partial_t {\bf E}_2 - \partial_x {\bf B} - {2 \over 3z}
{dz \over dy} {\bf E}_1 = 0 \\
\noalign{\vskip 14 pt}
& & J^0 = J^1 = {\gymb \over 4 z(y)} \sum_{I=1}^{16} \delta(y - y_I)
\langle \overline{\chi}_I \chi_I \rangle \,\,\, .
\eeas
When the fermions $\chi_I$ are in their ground state, they carry an
induced current if an electric field ${\bf E}_1$ is present, since
\[
\langle
0\vert \overline{\chi}_I \chi_I \vert 0 \rangle = - {1 \over 2 \pi}\,
{1 \over \partial_-} {\bf E}_1 \,.
\]
This can be shown by integrating out the $\chi_I$ fermions, as in
(\ref{AbelianEOM}).  We are interested in configurations with ${\bf
E}_1 = 0$.  Then this vacuum current vanishes, and $\langle 0\vert
\overline{\chi}_I \chi_I \vert 0 \rangle = 0$.

But now suppose $\chi_I$ has a zero mode, and consider the normalized
state $\vert 1 \rangle$ created by acting on the vacuum with this zero
mode.  There is a current present in this state, since $\langle 1\vert
\overline{\chi}_I \chi_I \vert 1 \rangle = {1 \over 2 \pi \tRi}$.  This
in turn implies that the electric and magnetic fields are
discontinuous at $y = y_I$:  ${\bf E}_2+{\bf B}$ jumps by ${\gymb \over
4 \pi \tRi z(y_I)}$, while ${\bf E}_2 - {\bf B}$ is continuous.  The
general static BPS solution to the equations of motion, taking these
discontinuities into account, is
\beas
{\bf E}_2(y) & = & {{\bf E}_0 + {\bf B}_0 \over 2 z^{5/3}(y)}
+ {{\bf E}_0 - {\bf B}_0 \over 2 z^{1/3}(y)} \\
\noalign{\vskip 6 pt}
{\bf B}(y) & = & {{\bf E}_0 + {\bf B}_0 \over 2 z^{5/3}(y)}
- {{\bf E}_0 - {\bf B}_0 \over 2 z^{1/3}(y)}
\eeas
where the integration `constants' ${\bf E}_0$, ${\bf B}_0$ are now
only piecewise constant.
\beas
{\bf E}_0 - {\bf B}_0 & = & \hbox{\rm const.} \\
{\bf E}_0 + {\bf B}_0 & = & \left\lbrace
\begin{array}{l}
\hbox{\rm piecewise constant, increases by ${\gymb z^{2/3}(y_I)
\over 4\pi\tRi}$} \\
\hbox{\rm at $y=y_I$ if a zero mode is excited}
\end{array}\right.
\eeas
One still has to impose the flux quantization conditions
(\ref{FluxQuant}) to completely fix ${\bf E}_0$ and ${\bf B}_0$.

The Hamiltonian is still given by
\[
H_{U(1)} = {1 \over \gymb} \int_{\widetilde{C}_2} d^2x \, {1 \over z(y)} \,
\left({({\bf E}_0 + {\bf B}_0)^2 \over z^{4/3}(y)} +
{({\bf E}_0 - {\bf B}_0)^2 \over z^{-4/3}(y)} \right)\,.
\]
So exciting $\chi_I$ zero modes will generically cost energy, in
accord with one's expectation of gauge symmetry breaking in the
heterotic string.

\subsection{Matching to Heterotic States}

The masses of the corresponding BPS states of the heterotic string are
well-known; see for example \cite{Ginsparg}.  We collect them here
since they provide a way to normalize the M(atrix) theory
parameters to the parameters of the heterotic string.
\begin{enumerate}
\item {\tt neutral momentum mode excitations}
\[
M^2 = \left(n \over R_H\right)^2
\]
\item {\tt neutral winding mode excitations}
\[
M^2 = \left({m R_H \over \alpha'_H}\right)^2 \left(1 + \half \alpha'_H
\vert A \vert^2\right)^2
\]
\item {\tt charged momentum mode excitations}
\[
M^2 = \left({n \over R_H} - A \cdot v\right)^2
\]
\item {\tt charged winding mode excitations}
\[
M^2 = \left({m R_H \over \alpha'_H} \left(1 + \half \alpha'_H \vert A \vert^2
\right) + A \cdot v \right)^2
\]
\end{enumerate}
The light-front energies of these states are related to their masses
by $E_{LF} = {R_{11} \over 2 N} M^2$.  In these expressions, $n$ and
$m$ are the momentum and winding quantum numbers, $A$ is the heterotic
Wilson line, and $v$ is a root of $E_8 \times E_8$.  The heterotic
Wilson line is normalized so that parallel
transport around ${\bf S}^1$ generates a phase $e^{i 2 \pi R_H A}$.
Recall that the roots of
$E_8$ are given by the roots of $SO(16)$ $\pm e_i \pm e_j$ together
with the weights of an $SO(16)$ spinor: $\pm \half e_1 \pm \cdots \pm
\half e_8$ with an even number of $(+)$ signs.

We need to relate the M(atrix) theory parameters $\lbrace \gymb, \tRi,
\tRii, y_I \rbrace$ to the heterotic parameters $\lbrace g_H,
\alpha'_H, R_H, A_I \rbrace$.  This can be done as follows.
\begin{itemize}
\item
A heterotic string wound on the regulator circle is identified with a
photon propagating around ${\bf S}^1$ in M(atrix) theory.  Equating
the energies of these states provides a relation ${ R_{11} \over
\alpha'_H} = {1 \over \tRi}$ which fixes $\tRi$.
\item
A neutral heterotic momentum mode excitation is identified with a quantum of
electric flux ${\bf E}_2$ in M(atrix) theory.  This can be used to
fix $\tRii$.
\item
A neutral heterotic winding mode excitation is identified with a quantum of
magnetic flux in M(atrix) theory.  This can be used to fix $\gymb$.
\item
Matching charged states of the heterotic string to states in M(atrix)
theory with $\chi_I$ zero modes excited fixes the D8-brane locations
$y_I$.
\end{itemize}
This procedure is closely related to the
matching of states that was carried out in
\cite{polchinskiwitten}.

\subsection{T-duality from S-duality}

T-duality of the heterotic string gets modified in the presence of
a Wilson line.  It acts according to \cite{Ginsparg}
\bea
\label{HeteroticTDuality}
R_H & \rightarrow & R_H^{\, \rm dual} = {\alpha'_H \over R_H}
\left(1 + \half \alpha'_H \vert A \vert^2 \right)^{-1} \\
A \,\,\,\,\, & \rightarrow & A^{\rm dual} = - A \,. \nonumber
\eea
This is clearly a symmetry of the BPS spectrum given above, with
momentum and winding exchanged.  It is expected to be a symmetry of
the full theory \cite{TDualityReview}.

As such, it should make an appearance in M(atrix) theory.  Given that
heterotic momentum is realized as electric flux, while heterotic
winding is realized as magnetic flux, we see that T-duality must be
realized as the interchange ${\bf E}_2 \leftrightarrow {\bf B}$ in
M(atrix) theory.  That is, heterotic T-duality must be realized in
M(atrix) theory as electric--magnetic S-duality in (2+1)-dimensions,
reminiscent of the way that T-duality of type \II~is realized in
M(atrix) theory compactified on ${\bf T}^3$ \cite{susskind},
\cite{ganorramgoolamtaylor}.

Usually, electric-magnetic duality is not possible in (2+1)
dimensions, because the number of independent components of the
electric and magnetic fields are not the same.  S-duality is made
possible in M(atrix) theory by the $\integer_2$ orbifold
projection, which eliminates states with electric flux ${\bf E}_1$.
We emphasize that this S-duality is unique to
heterotic M(atrix) theory.  For example in ${\bf T}^2$
compactifications of M-theory, the corresponding (2+1)-dimensional
M(atrix) gauge theory does not have S-duality.  This is expected,
since ${\bf T}^2$ compactification describes two inequivalent
string theories -- \IIA~and \IIB~-- that are mapped into each other by
T-duality.

Although we by no means have a proof that S-duality is a symmetry of
M(atrix) theory, some evidence is available from the BPS spectrum.
This is particularly clear from the spectrum of electric and magnetic
fluxes for unbroken $E_8 \times E_8$, given in (\ref{fluxes}).  The
spectrum, which is invariant under electric--magnetic duality, also
shows that, as one would expect, S-duality acts on the M(atrix) theory
by inverting the dimensionless coupling.

\section{Conclusions}

In this paper, we have begun to investigate the heterotic Wilson line
moduli of M(atrix) theory.  When the heterotic string is compactified
on a circle, the M(atrix) gauge theory lives on an orbifold ${\bf S}^1
\times ({\bf S}^1/ \integer_2)$.  A twisted sector of chiral
(1+1)-dimensional fermions must be introduced to cancel gauge
anomalies.  When $E_8 \times E_8$ is unbroken, these fermions are
located at the orbifold fixed points.
Turning on the heterotic Wilson
line moves these fermions into the interior of the orbifold.

The M(atrix) theory action is uniquely determined by gauge and
supersymmetry anomaly cancellation.  It involves position-dependent
Yang-Mills and Chern-Simons couplings, which explicitly break
(2+1)-dimensional Lorentz invariance.  From the string theory point of
view, this can be thought of as incorporating the massive
\IIA~supergravity background into M(atrix) theory.  This provides a
novel way in which a D2-brane can be regarded as a probe of background
geometry.  It also suggests that a full understanding of the moduli
space of M(atrix) theory will involve the study of non-Lorentz
invariant field theories.

The M(atrix) gauge theory has the appropriate BPS states, with the
correct gauge quantum numbers, to match the BPS states of the
heterotic string.  Identifying the BPS states serves to relate the
parameters of M(atrix) theory to the parameters of the heterotic
string.  But it also reveals a novel form of electric--magnetic
duality in (2+1)-dimensions, which corresponds to T-duality of the
heterotic string.  This S-duality arises in M(atrix) theory because
the gauge theory is defined on an orbifold ${\bf S}^1 \times ({\bf
S}^1/ \integer_2)$, and it is not possible to have electric flux along
the ${\bf S}^1$ direction.

Several directions for further investigation suggest themselves.
Although it is qualitatively possible to understand how the BPS
spectrum of the M(atrix) theory is deformed when the heterotic Wilson
line is turned on, a more precise understanding of the BPS states of
M(atrix) theory is clearly desirable.  One would also like to have a
better understanding of the global structure of the Wilson line moduli
space in M(atrix) theory.  Being proposed as the definition of
M-theory, M(atrix) theory should be able to describe the full Narain
moduli space, perhaps in a more elegant manner than string theory
itself.  It would be very interesting, and would perhaps shed light on
these matters, to relate the M(atrix) theory described in this paper
to the proposal~\cite{Govindarajan, BerkoozRozali} for the heterotic
string compactified on ${\bf T}^3$.

\vskip 1.0 cm
\centerline{\bf Acknowledgments}
We thank E.~D'Hoker, M.~Porrati, L.~Susskind, F.~Wilczek and E.~Witten
 for valuable discussions.

\appendix
\section{Supersymmetry in 2+1 Dimensions}
We briefly review the construction of supersymmetric actions in 2+1
dimensions in terms of ${\cal N} = 1$ superfields \cite{Superspace,
AffleckHarveyWitten}.  Our metric is $(-++)$, $\epsilon^{012} = +1$,
and we use the spinor conventions given in section 2.1.

Superspace in 2+1 dimensions has, besides the usual bosonic
coordinates $x^\alpha$, a single Majorana spinor coordinate
$\theta_a$.  We sometimes denote $\theta^a \equiv \left(\theta^T
\gamma^0\right)_a$.  The supercharge and supercovariant derivative are
\beas
\hbox{\cmss Q}_a & = & {\partial \over \partial \theta^a} + i
\left(\gamma^\alpha \theta\right)_a  \partial_\alpha \\
\hbox{\cmss D}_a & = & {\partial \over \partial \theta^a} - i
\left(\gamma^\alpha \theta\right)_a \partial_\alpha
\,\,\, .
\eeas

The scalar multiplet contains a real scalar field $X$, a Majorana
fermion $\psi$, and a real auxiliary field $F$.  It is described by an
unconstrained real scalar superfield $\Phi$, with component expansion
\[
\Phi = X + \bar{\theta} \psi + \half \bar{\theta} \theta F
\,\,\, .
\]
In components, the supersymmetry transformation $\delta_\epsilon \Phi
= \bar{\epsilon} \, \hbox{\cmss Q} \Phi$ reads
\beas
\delta_\epsilon X & = & \bar{\epsilon} \psi \\
\delta_\epsilon \psi & = & - i \gamma^\alpha \partial_\alpha X \epsilon
+ F
\epsilon \\
\delta_\epsilon F & = & - i \bar{\epsilon} \gamma^\alpha \partial_\alpha
\psi
\,\,\, .
\eeas

The vector multiplet includes a gauge field $A_\alpha$ and a Majorana
fermion $\lambda$.  It is described by a real spinor superfield
$\Gamma_a$, with component expansion (in Wess-Zumino gauge)
\[
\Gamma_a = i A_\alpha \left(\gamma^\alpha \theta\right)_a +
\bar{\theta} \theta \lambda_a \,\,\, .
\]
Supersymmetry acts by
\beas
\delta_\epsilon A_\alpha & = & - i \bar{\epsilon} \gamma_\alpha
\lambda \\
\delta_\epsilon \lambda & = & \half F_{\alpha\beta} \gamma^{\alpha\beta}
\epsilon
\,\,\, .
\eeas

The gauge-covariant derivative $\nabla_a = \hbox{\cmss D}_a - i \Gamma_a$
lets us write a kinetic term for the scalar multiplet: ${\rm Tr} \,
\left(\nabla^a \Phi \nabla_a \Phi\right)\vert_{\tbt}$.  The field
strength $W_a$ is given by
\beas
W_a & = & - \hbox{\cmss D}^b \hbox{\cmss D}_a \Gamma_b + i [\Gamma^b,
\hbox{\cmss D}_b \Gamma_a] + {1 \over 3}[\Gamma^b,\lbrace\Gamma_b,
\Gamma_a\rbrace] \\
& = & 2 \lambda_a + F_{\alpha\beta} \left(\gamma^{\alpha\beta}
\theta \right)_a - i \bar{\theta} \theta \left(\gamma^\alpha
D_\alpha \lambda\right)_a \,\,\, .
\eeas
The usual supersymmetric Yang-Mills term is ${\rm Tr}
\, \left(W^a W_a\right)\vert_{\tbt}$.  In (2+1)-dimensions, one also
has the option of introducing a Chern-Simons term for the gauge field.
The supersymmetric Chern-Simons term is given by the top component of a
(non-gauge-invariant) superfield ${\cal G}$:
\beas
{\cal G} & \equiv & \half {\rm Tr} \, \left[ \Gamma^a W_a + {i \over 6}
\left\lbrace\Gamma^a,\Gamma^b\right\rbrace \hbox{\cmss D}_b\Gamma_a
+ {1 \over 12} \left\lbrace\Gamma^a,\Gamma^b\right\rbrace
\left\lbrace\Gamma_a,\Gamma_b\right\rbrace \right] \\
{\cal G}\vert_{\tbt} & = & {\rm Tr} \left[
\epsilon^{\alpha\beta\gamma} \left(A_\alpha \partial_\beta A_\gamma
+ i {2 \over 3} A_\alpha A_\beta A_\gamma \right)
+ \overline{\lambda} \lambda \right]
\,\,\, .
\eeas


\end{document}